\documentclass{article}
\usepackage[english]{babel}
\usepackage[T2A]{fontenc}
\usepackage[utf8]{inputenc}
\usepackage{mathtext}
\usepackage{graphicx}
\usepackage{amsfonts}
\usepackage{amssymb}
\usepackage{amsmath}
\usepackage{float}
\usepackage{indentfirst}
\usepackage[a4paper,margin=2.5cm]{geometry}
\begin{document}
\title{Conductivity measurements in JCH like models}

\author{Yuri I. Ozhigov\thanks{Lomonosov Moscow State University, Faculty of Computational athematics and Cybernetics, Moscow, Russia}\hspace{0.15cm} and Nikita A. Skovoroda\thanks{Lomonosov Moscow State University, Faculty of Computational Mathematics and Cybernetics, Moscow, Russia}
}

\maketitle

\begin{abstract}
We consider the conductivity of excitations in short chains of optical cavities with two-level atoms in the models of JCH type, where either we explicitly take into account the photon jumps between atoms, or is merely a transfer of excitation from atom to atom.  We found a non-trivial dependence of the conductivity on the intensity of runoff and inflow (quantum bottleneck) in the presence of dephasing noise (the effect of dephasing assisted transport).

\vspace{0.2cm}

\end{abstract}


\section{Introduction and background}

Complex systems is the strongest challenge to quantum theory since its inception. It is related to the fundamental problem of the complexity of the description of many-particle ensembles at the quantum level, which led to the idea of a quantum computer. The most critical element in our understanding of the physics of quantum computing is decoherence, which can be described within the concept of an open quantum system (see \cite{BPe}).

Kossakowsky-Lindblad master equation (see \cite{Kos} and \cite{Lin}), gives the simplest description of an open quantum system, where the influence of environment is expressed by permanent measurements of the system and is represented as Markov random process. Despite the fact that in many cases this influence can hardly be considered as Markov process, and there are non Markovian approaches (see., Eg, \cite{HL}; the analytical analysis of interaction with invironment can be found in \cite{11}), the dynamics in terms of the master equation has the simplest form, it makes it easy to compare result with the unitary dynamics of the expansion of the system, where we include some part of the environment to the considered system.

Coherence means the presence of off-diagonal elements of the density matrix; quantum effects are always associated with coherence, and the effects of information nature as quantum algorithms, moreover, require even the existence of entanglement. Influence of environment usually leads to the suppression of coherence and treated as noise in the practical use of quantum effects (see \cite{BPe}). However, in some cases, on the contrary, the influence of the environment can help to maintain the desired quantum effects. The most famous example - quantum Zeno effect, in which the measurements allow to conserve the desired quantum state (see the work \cite{HS} Zeno effect in  NMR- quantum computing as the freezing of quantum states, coherence protection by Zeno effect in \cite{BAC}). It is also possible to suppress decoherence by a special noise (see, for example \cite{ST}). In the work \cite{MP}the  non-trivial relationship between the unitary dynamics with Lindblad master equation is described, in particular, the possibility of noise smoothing by a unitary control.

The influence of noise is very important for the conductivity of excitations in groups of two-level atoms enclosed in the optical cavities with the possibility of exchange of photons between the cavities. Excitation conductivity can be used to analyze the transmission of energy and information at the quantum level, in artificial devices and living organisms (see, for example \cite{4} , \cite{5} , \cite{HP}) where the addition of the noise to the model makes it more realistic. 

We can formulate the conductivity of excitations in terms of quantum walks (\cite{Am}, \cite{RNB}), which have a  valuable difference from classical random walks. For example, quantum conductivity quadratically exceeds classical: the conductivity of the classical walk on a linear chain takes time of the order of  squared length of the chain, whereas quantum - of the order of the length ( \cite{Am}, \cite{Aga},  \cite{CFG}). For the walks on glued trees quantum speedup of the conductivity will be even exponential(see \cite{CFG}). 

We will consider two important quantum effect associated with conductivity: dephasing assisted transport (DAT) and quantum bottle-neck. DAT plays an important role in the light harvesting  Fenna-Matthews-Olson (FMO) complex in green sulfur bacteria (see \cite{FMO}).  The essence of the DAT effect is that the presence of noise at the resonant frequency increases transfer of excitations, so that there is an optimal intensity of the noise when the conductivity is maximal. In the works \cite{CC},\cite{HP},\cite{5},\cite{7},\cite{10},\cite{20}) DAT effect was investigated on the model of excitation transfer, without the explicit consideration of the interaction between photons and atoms. 

However, artificial reproduction of excitation transfer is important, for example, for the effective conversion of light into chemical energy by optical elements. To simulate such systems we must  take into account the interaction of atoms with light explicitly, as is done in this paper.

Quantum bottle-neck effect is that the maximum conductivity occurs at the optimum value the force of interaction of the end node with the sink (flow rate). The excess of this value does not lead to an increase in conductivity (as in the classical case where the conductivity increases monotonically with the intensity of runoff), but decreases it, until the complete disappearance. This is like DAT, purely quantum, and, moreover, counter-intuitive effect. It is noted in the paper \cite {10} (p. 10) that the bottle-neck can exist only in the case of coherent exchange between nodes, i.e. it has quantum nature, but there were no exact boundaries for Lindblad dephasing, in which this effect arises. In this paper we present such boundaries.

The physics of quantum effects in the conductivity of excitations is extremely important in biology, as well as in quantum technology, so it is of interest to study these effects in more detail in the framework of Jaynes-Cummings-Hubbard model (JCH), introduced in \cite{JC} , \cite{Hub}.  The explicit consideration of photons and their interaction with atomic excitations explains many quantum effects important for the applications.
 These effects can be used in quantum computing: optical CNOT gates (\cite{A}) , quantum computations on optical gates (see \cite{KM}, and \cite{KLM}) - quantum computations on linear optical gates and measurements, photonic quantum memory (see, for example, \cite{Moi}), the description of the other quantum effects and their possible applications can be found in \cite{RW} , \cite{13} , \cite{14} , \cite{15} , \cite{16} , \cite{17} , \cite{18}, \cite{8}.

Excitation conductivity has more detailed description in JCH model, which depends on its parameters: amplitude of photon jumps between cavities and amplitude of atom-photon interaction inside of one cavity. For example, in the work \cite{MC} excitation conductivity is considered in JCH model without phonons. We make such analysis for the model with phonons and establish the influence of atom-photon interaction to DAT effect. 

Our goal - the quantitative description of two quantum effects arising in transfer of energy along the chain of atoms concluded in optical cavities: quantum bottleneck and dephasing assisted transport (DAT). We formulate a model in which it is possible to combine the simultaneous influence of these effects on the excitation conductivity, and then present the results of numerical simulations.

We also give a brief qualitative explanation of these two effects, and indicate possible ways to use these effects to improve the quantum models of optical systems associated with the conversion of light energy into chemical.

\section{Quantum effects in the energy transfer}

We give a definition, starting with the most detailed model "excitations, photons, phonons." We consider a linear chain consisting of identical optical cavities with two level atoms inside. Each cavity  holds the photons with frequency $\omega_c$. Fock state with $n$ photons inside $i$-th cavity we denote by 
$|n\rangle_{fi_i}$. Inside each $i$-th cavity there is one two-level atom, which eigenstates: ground and excited we denote by $|0\rangle_{at\ i},\ |1\rangle_{at\ i}$ correspondingly. The difference between frequencies of these states (detuning) $d=\omega_c-\omega_a$ is small in comparizon with each of them: $d\ll \omega_c$ that allows to use rotating wave approximation for the atom-photon interaction at the large enough time frame, and Jaynes -Cummings model (JC). In addition, each atom is placed in a bath of thermal phonons that have dephasing effect on its excitation. In this paper all phonons have the same frequency $\omega_p$, for which DAT effect is maximal. In fact, this frequency is equal to the difference between eigen frequencies of the Hamiltonian of JCH model (see below) that are responsible for the transfer of excitations (see the work \cite{HP}). 

The transfer of energy from the cavity to cavity occurs through the flight of photons between the cavities where $-\frac{1}{h}\delta_{i, j} $ is the amplitude of the flight from the cavity $ j $ to the cavity $ i $ per unit of time, so that the amplitudes of the opposing flights are mutually conjugate, that is $\delta_{i,j}=\bar\delta_{j,i}$.

The Hamiltonian of our model is thus obtained by adding to the Jaynes-Cummings Hubbard Hamiltonian  $ H_{JCH} $ the term of exciton-phonon interaction  $ H_{int \ ep} $:
\begin{equation}
\begin{array}{ll}
H=&H_{JCH}+H_{int\ ep},\\ &H_{JCH}=h\omega_a\sum\limits_i\sigma^+_i\sigma_i+h\omega_c\sum\limits_ia^+_ia_i+\sum\limits_i
(\gamma a_i^+\sigma_i+\bar\gamma a_i\sigma_i^+)+\sum_{i\neq j}(\delta_{i, j}a_i^+a_j+\bar\delta{j,i}a_j^+a_i),\\
&H_{int\ ep}=g\sum\limits_i(b_i^++b_i)\sigma_i^+\sigma_i.
\end{array}
\label{H}
\end{equation}
Here $a_i^+,a_i$ denote operators of creation and annihilation of a photon in $i$-th cavity, $b_i^+,b_i^+$- denote creation and annihilation of a phonon in $i$-th cavity, $\sigma_i^+,\sigma_i$- denote creation and annihilation of the excited state of the corresponding atom (excitation),  $\gamma$ is the amplitude of photon emitting by excited atom in the unit of time. We assume that the photon jump is possible between the neighboring cavities only where it occurs with the same amplitude so that all $\delta_{i,j}=0$ for $|i-j|\neq 1$, and for $|i-j|= 1$, $\delta_{i,j}=\delta$. Constant $g$ is the intensity of exciton-phonon interactions (square root of Huang-Phys factors).

In the simplified model, which we call ''excitation-phonon'' photons are ignored, and excitations are transmitted from one atom to the other; its Hamiltonian $H_{ep}$ has the form  
\begin{equation}
H_{ep}=H_e+H_{int\ ep},\ H_e=h\omega_a\sum\limits_i\sigma^+_i\sigma_i+\sum_{i\neq j}(\mu_{i, j}\sigma_i^+\sigma_j+\bar\mu_{j,i}\sigma_j^+\sigma_i),
\label{ex}
\end{equation}
 The inflow of the energy can be viewed either as a constant occurrence of the excitation on the first node, taking the energy income from the outer bath, or simply set the initial state of the first atom as excited, regardless of photons. In the first case the bath inflows can create excitation irreversibly, or we can simply consider the initial state of a strongly excited field in the first cavity, which interacts with the first atom.

There are two ways for the simulation of dephasing as well. Either we consider it as the interaction with the explicit phonons, or as the irreversible process with only excitations and without explicit phonons. 

Combining all of the above methods of studying the conductivity, we get all kinds of particular computing models. In any case, for the irreversible models master equation of Kossakowsky-Lindblad should be applied:

\begin{equation}
\rho_{t + \delta t} = U_{\delta t}^*\rho_{t}U_{\delta t}
 + \delta t \sum\limits_{i}(L_i \rho L_i^* - \frac{1}{2}(L_i^* L_i \rho + \rho L_i^* L_i)),\ \ \ 
U_{\delta t} = e^{-\dfrac{i \cdot \delta t}{\hbar} H}.
\label{modeleq}
\end{equation}
 
 which contains the unitary dynamics $U_{\delta t}$ as the particular case. Here Lindblad operators $L_i$ describe the irreversible part of the process. For example, the irreversible runoff of excitations to the sink from the last node (end) is expressed by the operator $L_{sink}=g_{sink}|0\rangle_{end}|1\rangle_{sink}\langle 0|_{sink}\langle 1|_{end}$, where the positive coefficient  $g_{sink}$ expresses the intensity of runoff, dephasing in $i$-th node resulted from the interaction with the implicit phonons is reflected by the operator $L_{deph,\ i}=\sigma_i^+\sigma_i$, photon leak through the walls of the $i$-th cavity - by $L_{det,\ i}=a_i$, etc. 

We consider the resulting quantum effects of excitation conductivity.
We determine the conductivity by the degree of filling sink in a certain time.

1). Quantum bottleneck. Reduction of conductivity with increasing intensity of the runoff.
Graph of conductivity depending on the rate of the runoff is not monotonically increasing as would be in the case of classical conductivity, but has a local maximum at a finite rate $ g_{sink}^0 $. The nature of this counter-intuitive effect is purely quantum.

It can be explained by considering the dynamics of the population of the initial state according to the time when the runoff intensity is large. If there is no runoff, the dynamics will be similar to the graph of the cosine, which has a maximum point at the initial time. The runoff constantly destroys the excitation at the end of the chain that makes the density matrix close to the initial state in which a reduction of the excitonic population at the first atom decreases with time very weak. This process competes with the obvious classic excitonic population decrease with increasing runoff.

If the runoff is strong enough, it constantly holds the density matrix (and hence population of excitations in the first atom) in the initial state in which there is practically no reduction of excitonic population at all that results in '' freezing '' of the population, like in quantum Zeno effect, in which the effect of '' freezing '' comes from frequent measurements of the quantum state.
In the case of the classical conductivity instead of cosine fall of excitonic population we would have a linear decrease, and there would be no bottleneck. 

2). Dephasing assisted transport - DAT. A plot of the conductivity on the intensity of the noise (the number of phonons of the resonant frequency, or intensity of the  interaction between excitations and phonons, expressed in coefficients of Lindblad for dephasing) has a local maximum at a non-zero point. This means that there is some non-zero noise intensity at which the conductivity is maximal. There are quantum processes, connected with the conductivity, the effectiveness of which is enhanced with the increase of noise, for example, mixing at random walks on graphs (\cite{FST}). Effect of DAT stands out among them due a special role it plays in biology.

DAT as a purely quantum effect. Dephasing is direct suppression of the off-diagonal elements of the density matrix. Quantum dynamics is obtained by adding the amplitude states, the evolution of which leads to the same point of the classical space. Dephasing shifts the phase of the amplitude to a certain area. If this area would be narrower than the natural spread of the phase of states in the absence of dephasing, this influence will make the interference more constructive, thus increasing conductivity. This effect can be illustrated by the example of road traffic. If possibilities of cars are the same and allow rapid acceleration and braking, then synchronous mode of motion (phase conservation) will support high bandwidth of the road. But if the possibilities of vehicles are very different, then to increase bandwidth it is needed to the contrary, smooth mode of braking and acceleration, i.e. dephasing.

 \begin{figure}
  \begin{center}
   \includegraphics[width=0.45\textwidth]{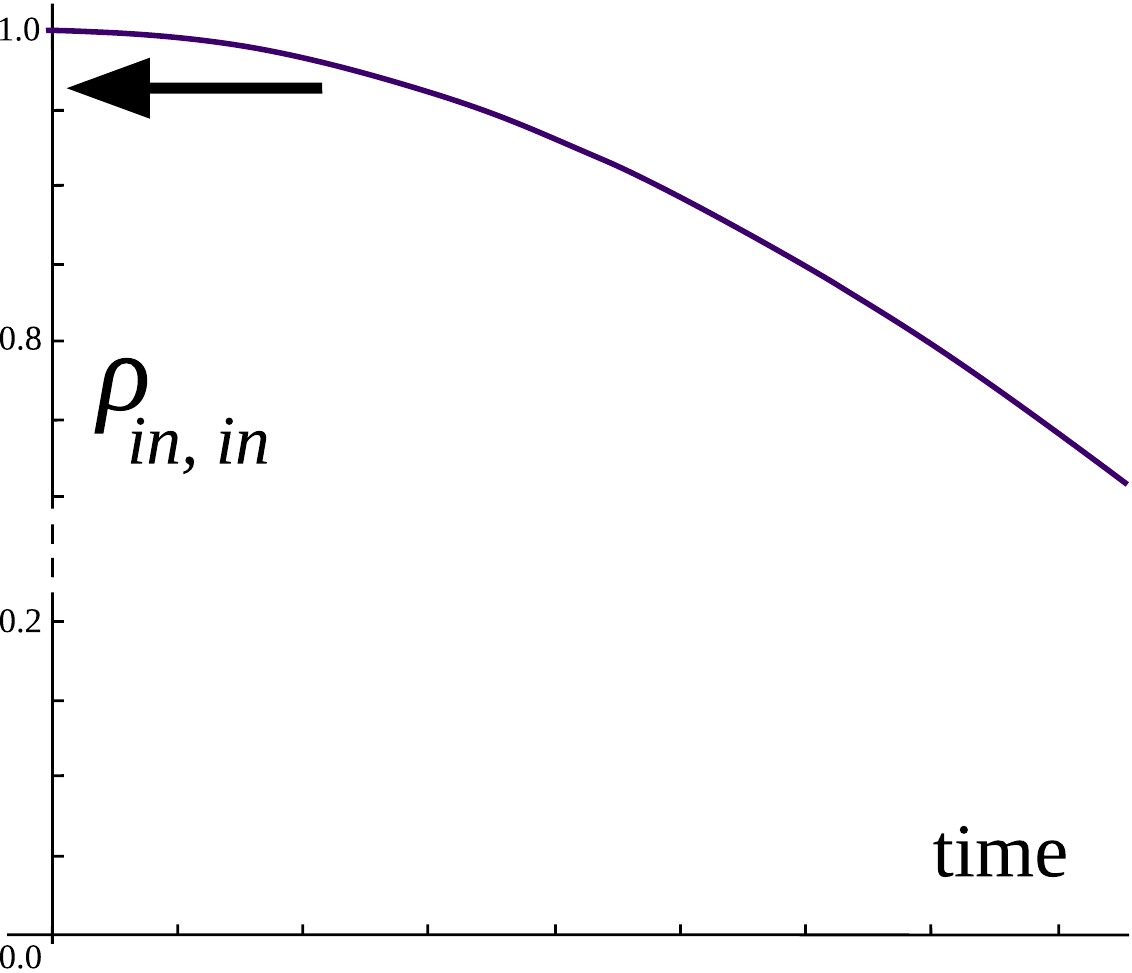}
  \end{center}
 \caption{Dynamics of the population of the initial state with a large runoff intensity. Runoff shifts the density matrix to the initial state, the population of excitation in the first atom (for two-atom chain this is $ \rho_ {in, in} $) is shifted to the maximum point, and remains almost constant for a long time, preventing the sink filling.}
\end{figure}

\pagebreak
\section{Qubit-based model}

\subsection{Evolution equations}
Kossakowski--Lindlbad equation in diagonal form is used for calculating non-unitary evalution:
\begin{equation}
i\hbar\frac{\partial\rho}{\partial t}=[H,\rho ]+i\sum\limits_{i}^{N^2-1}\gamma_i(A_i\rho A^*_i-\frac{1}{2}(A^*_iA_i\rho+\rho A^*_iA_i))
\label{lindblad}
\end{equation}


Unitary evalution is optimized by diagonalizing the Hamiltonian.

In both cases, the evolution operator is determined by a Hamiltonian $H$ with a set of non-normalized matrices $L_i = \sqrt{\gamma_i} A_i$. 


For general (non-unitary) case, the evolution is computed by the equations (\ref{modeleq}) directy.


These evolution equations, as well as the measurement implementation, are independent of the used model. Various models could be created by constructing a Hamiltonian $H$ and a set of Lindblad matrices $L_i$ according to that model rules and semantics.

\subsection{Qubit-based model construction}
In order to allow simple operator-based constructing of Hamiltonians and Lindblad matrices we build the models by binding one a several qubits (a qubit group) to a specific state parameter, for example to the excitation of the first atom (see \cite{3}). The model construction methods work with virtually huge matrices: for example for a chain of one-levelled excitons of a length $100$, it would have a virtual Hamiltonian of size $2^{100} * 2^{100}$ (one qubit per each exciton). If excitations were four-leveled, that would make two qubits for each excitation. These matrices are not actually constructed in memory, an energy-limiting projection is used which selects only those states that have the energy between the specified lower and upper bounds. The resulting Hamiltonian and Lindblad matrices which are used in the actual calculation have much smaller sizes due to this projection.

\subsection{A chain of excitation-photon-phonon cavities}

A general chain model was constructed. This model implements a chain of cavities, each of which has simular multi-level excitation, photon field, and phonons.
It has energy limiting support and supports attaching an input source and an output sink.

\begin{figure}[H]
\begin{center}
\includegraphics[width=0.7\textwidth]{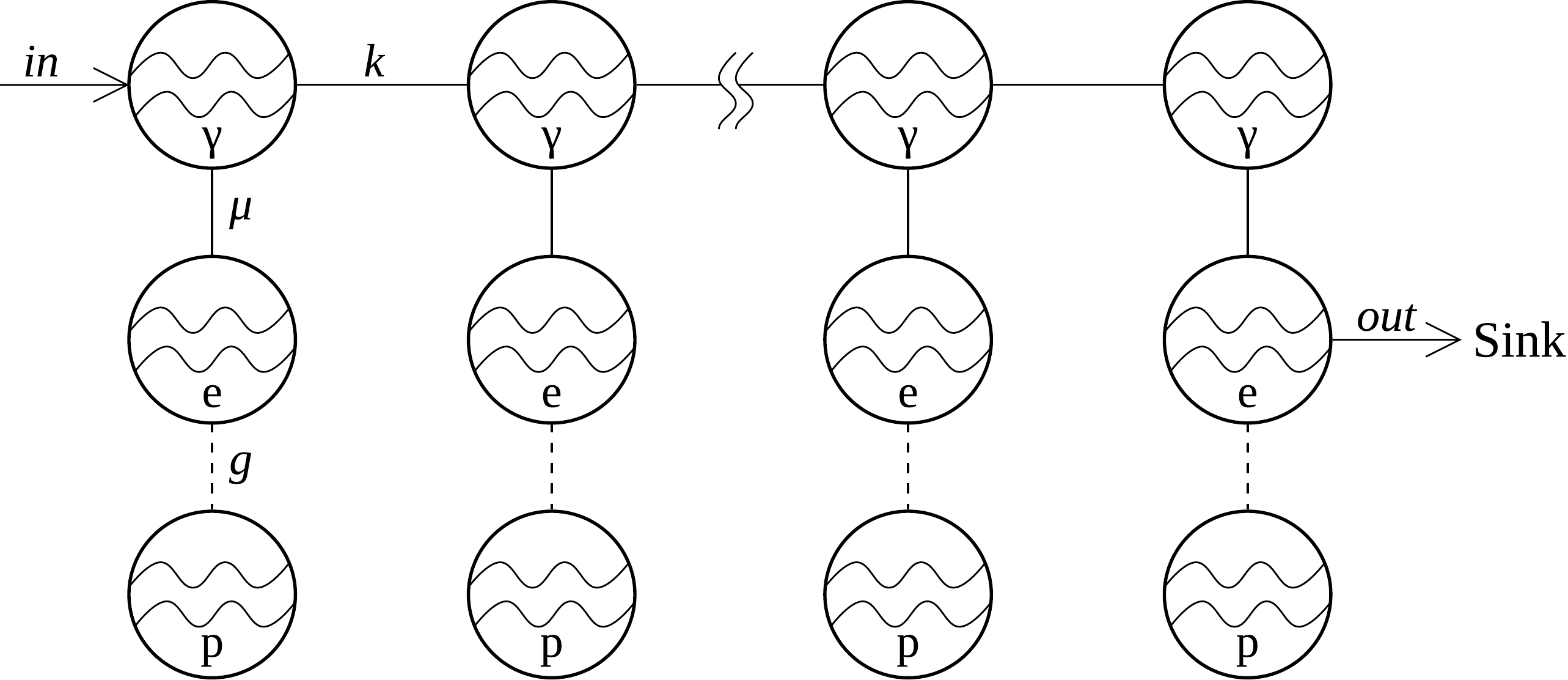}
\end{center}
\vspace{0.2cm}
   \caption{ \label{fig:model}  The exciton-photon-phonon cavities chain schematics. }
\end{figure}

There are three types of qubit groups in the chain:
\begin{itemize}
 \item $b_i$ is the phonon state for atom $i$, \: 1 --- the corresponding phonon is present, 0 --- the corresponding phonon is absent;
 \item $a_i$ is the exciton state for atom $i$, \: 1 --- excited state, 0 --- not excited;
 \item $p_i$ is the photon state for atom $i$, \: 1 --- there is a photon, 0 --- no photon.
\end{itemize}

The chain model has the following parameters:
\begin{itemize}
 \item $N_{atoms}$ --- the number of atoms,
 \item $k$ --- photon tunnelling rate between neighbouring cavities,
 \item $\mu$ --- photon--atom interaction strength,
 \item $g$ --- phonon--atom interaction strength,
 \item $\omega_a$ --- frequency of one excitation (atomic transition frequency),
 \item $\omega_p$ --- frequency of one photon (cavity frequency),
 \item $\omega_g$ --- frequency of one phonon,
 \item $in$ --- input rate, replenishment coefficient for the first excitation $a_1$,
 \item $out$ --- output rate, sink coefficient for transferring excitation from $a_{N_{atoms}}$ to $s$.
\end{itemize}

Hamiltonian of the system equals to Jaynes--Cummings--Hubbard Hamiltonian:
\begin{equation}
\begin{aligned}
H = & \sum\limits_i\omega_pp^+_ip^-_i +
          \sum\limits_i\omega_aa^+_ia^-_i +
          \sum\limits_i\omega_bb^+_ib^-_i +
          \sum\limits_i(kp^+_{i+1}p^-_i + k^*p^+_ip^-_{i+1})\ + \\
        & \sum\limits_i(\mu p^-_ia^+_i + \mu^* p^+_ia^-_i) +
          (g + g^*)\sum\limits_i((b^-_i + b^+_i)a^+_ia^-_i).
\end{aligned}
\label{jch}
\end{equation}

The sink $s$ is attached to the last exciton in the chain. Input and output is performed using Lindblad operators:

\begin{equation}
\begin{aligned}
L_{in} &= in * p^+_1 \\
L_{out} &= out * s^+p^-_{N_{atoms}} \\
\end{aligned}
\label{jchlr}
\end{equation}

In all the follow-up graphs and results, it is taken $\omega_a = \omega_p = 0.1$ and $\omega_g = 0.01$.

\subsection{Dephasing models}

Two dephasing models were implemented:
\begin{itemize}
 \item Unitary-based: with explicit phonons as shown on \ref{fig:model}. \\ Hamiltonian of the system has the $(g + g^*)\sum\limits_i((b^-_i + b^+_i)a^+_ia^-_i)$ part.
 \item Lindblad-based: there are no explicit photons in the system. Hamiltonian of the system does not have the corresponding part, Lindblad-like dephasing operators $D_{i} = g * p^+_ip^-_i$ are used instead\cite{5}.
\end{itemize}

With the addition of Lindblad-like dephasing operators, evaluation equation \ref{modeleq} takes the following form:
\begin{equation}
\rho_{t + \delta t} = U_{\delta t}^*\rho_{t}U_{\delta t}
 + \delta t \sum\limits_{i}(L_i \rho L_i^* - \frac{1}{2}(L_i^* L_i \rho + \rho L_i^* L_i))
 + \delta t \sum\limits_{i}(D_i^* D_i \rho D_i^* D_i - \frac{1}{2}(D_i^* D_i \rho + \rho D_i^* D_i))
\end{equation}

\section{Quantum bottleneck}
\label{bottleneck}

\begin{figure}[H]
\begin{tabular}{c c c}
\includegraphics[width=0.45\textwidth]{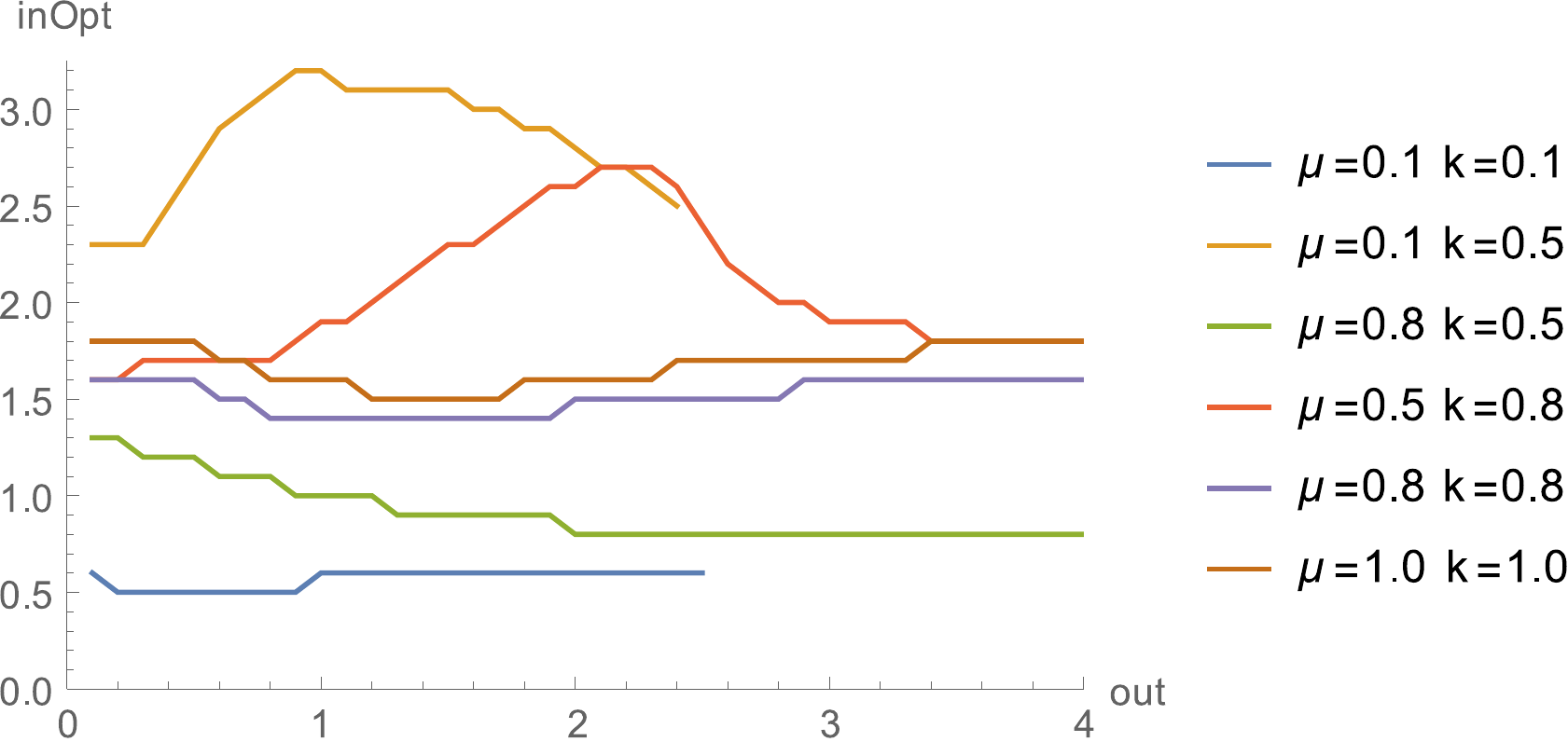} & & \includegraphics[width=0.45\textwidth]{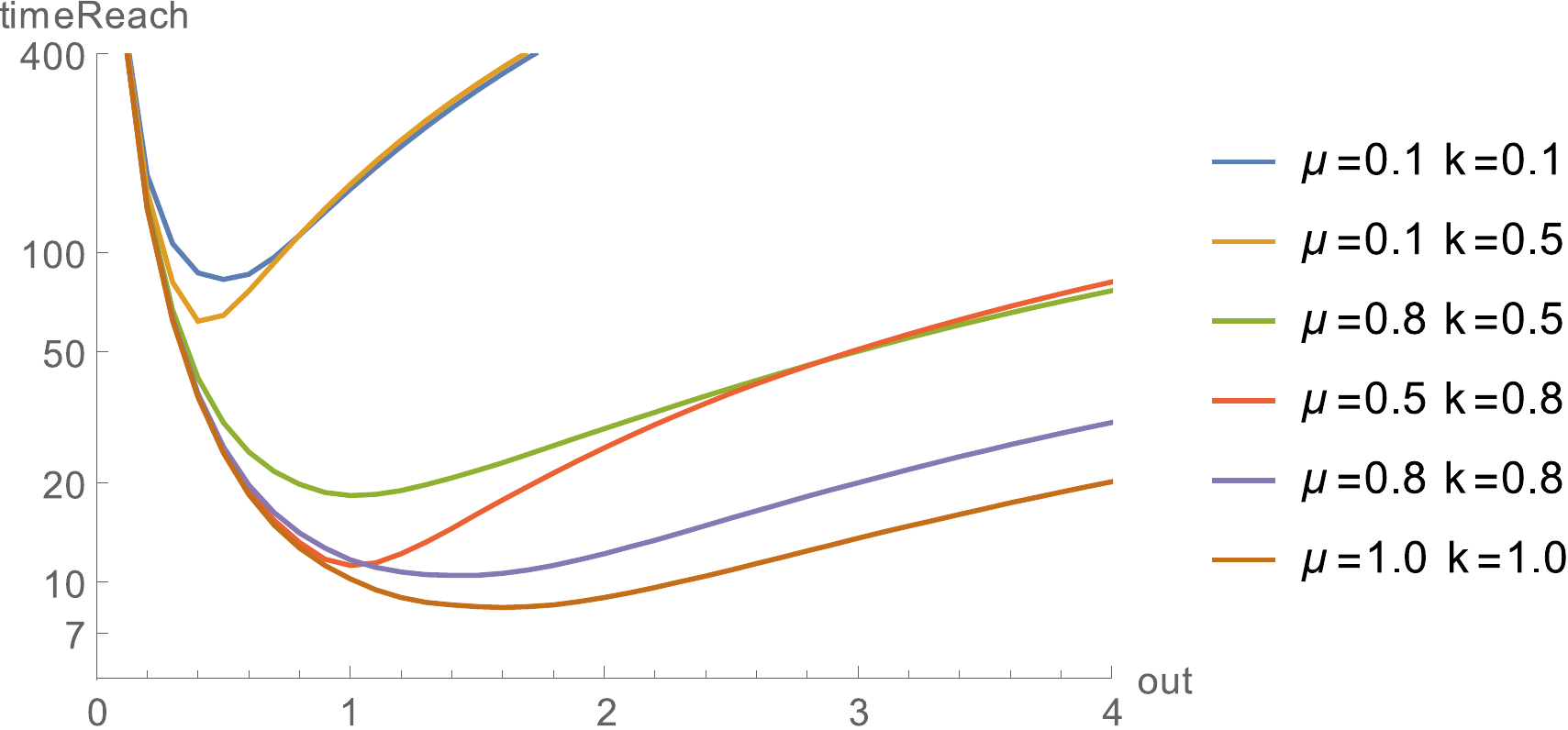} \\
The optimal input rate, over the output rate & & Time taken to reach target sink value $0.995$ \\
& & using the optimal input rate, over the output rate \\
\end{tabular}
\vspace{0.2cm}
   \caption{ \label{fig:2-out-optimal-in} Optimal input rate for for a chain of two atoms. }
\end{figure}

Graph \ref{fig:2-out-optimal-in} shows the optimal value of the input rate depending on the output rate for an chain of two cavities with one excitation and one possible photon each, for sample $k$ and $\mu$ values. The second part of the graph shows the time taken to reach target sink value using the optimal input rate for the selected output rate. It could be seen that even with optimizing the input rate, the quantum bottleneck effect is observed. Graphs are capped at $timeReach = 400$, this is the reason why the first part of the graph is incomplete for $\mu = 0.1$.

\begin{figure}[H]
\begin{tabular}{c c c}
\includegraphics[width=0.45\textwidth]{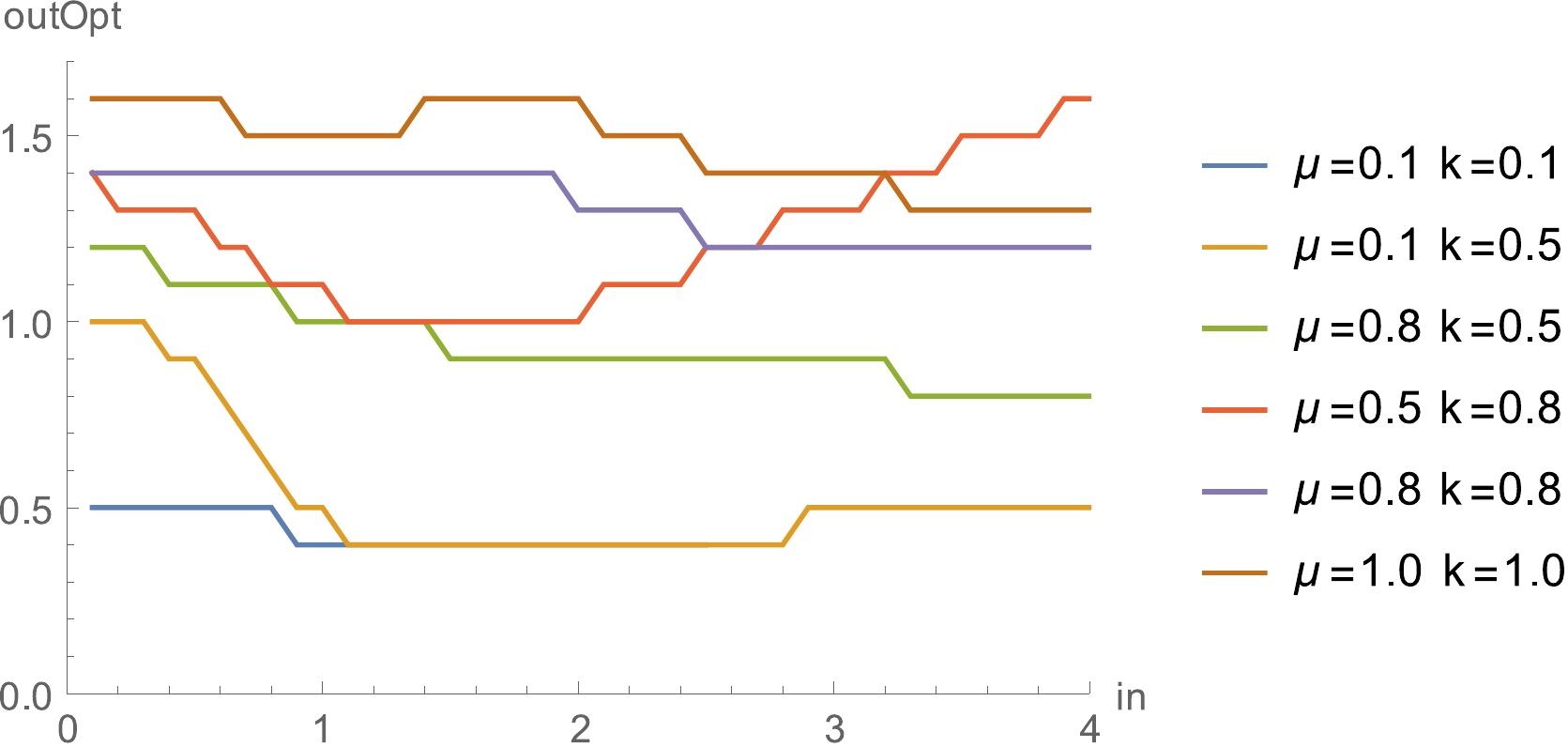} & & \includegraphics[width=0.45\textwidth]{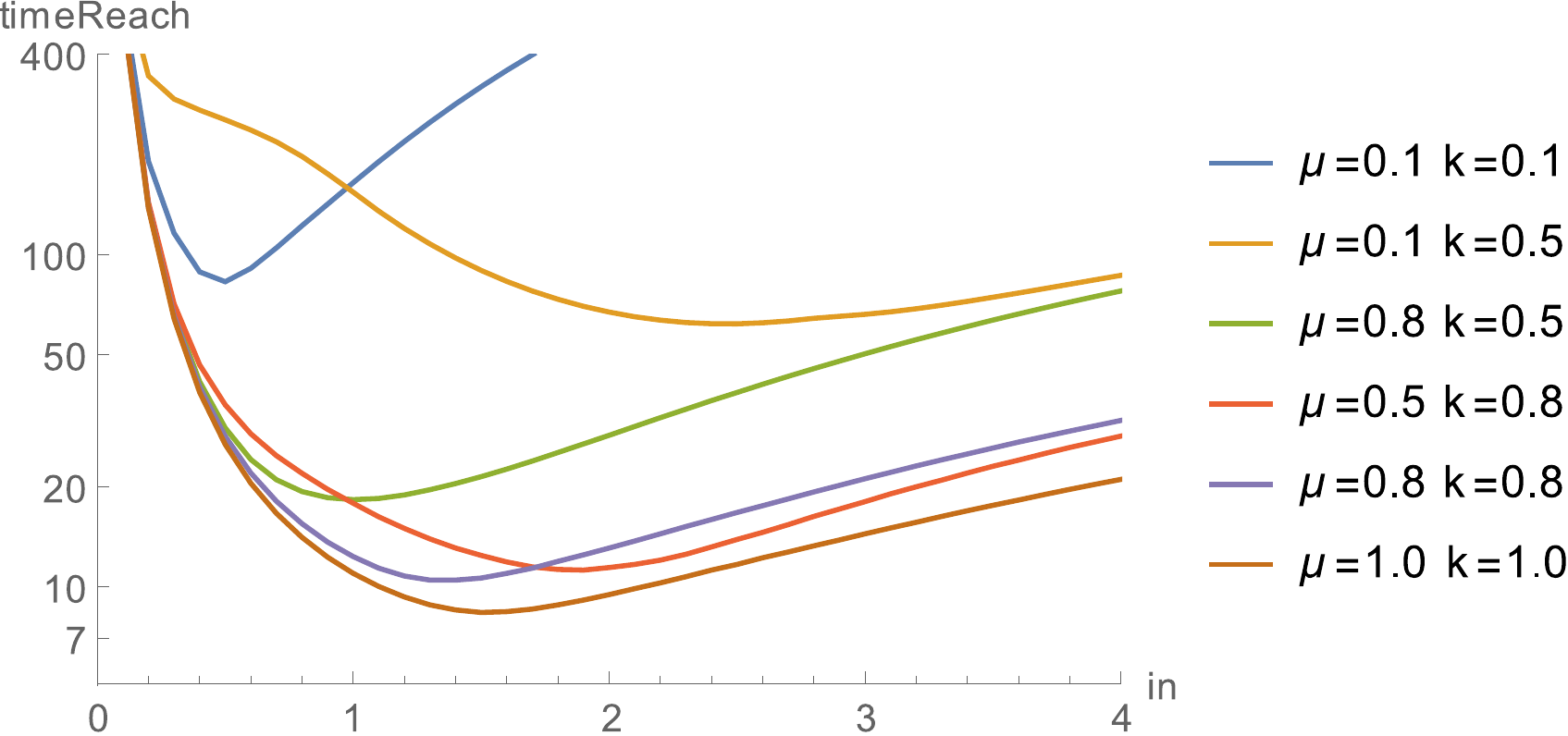} \\
The optimal output rate, over the input rate& & Time taken to reach target sink value $0.995$ \\
 & & using the optimal output rate, over the input rate \\
\end{tabular}
\vspace{0.2cm}
   \caption{ \label{fig:2-in-optimal-out} Optimal output rate for for a chain of two atoms. }
\end{figure}

Graph \ref{fig:2-in-optimal-out} shows the optimal value of the output rate depending on the input rate for the same chain of two cavities.

\vspace{-0.2cm}
\begin{figure}[H]
\begin{center}
\includegraphics[width=0.5\textwidth]{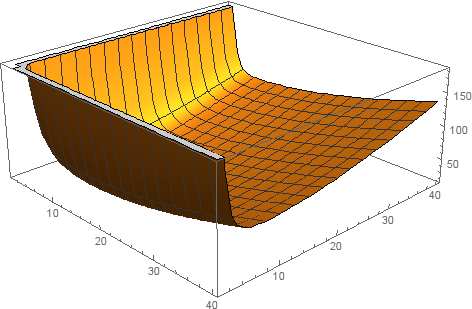} \\
\end{center}
   \caption{ \label{fig:2-inout-3d} The dependency of time taken to reach target sink value $0.995$ over $in$ and $out$ rates. $\mu$ = 0.8, $k$ = 0.5}
\end{figure}

\begin{figure}[H]
\begin{tabular}{c c c}
\includegraphics[width=0.45\textwidth]{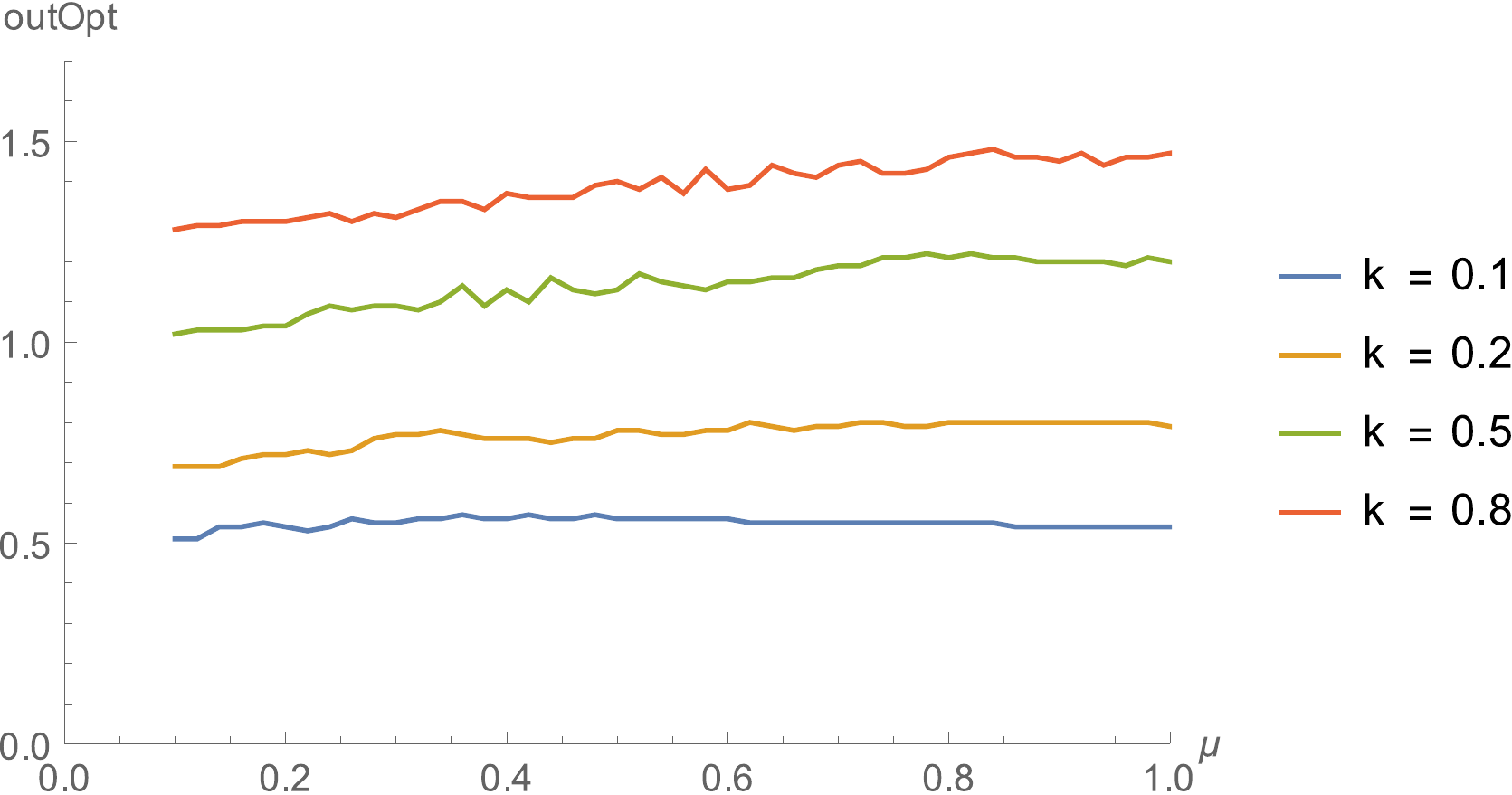} & & \includegraphics[width=0.45\textwidth]{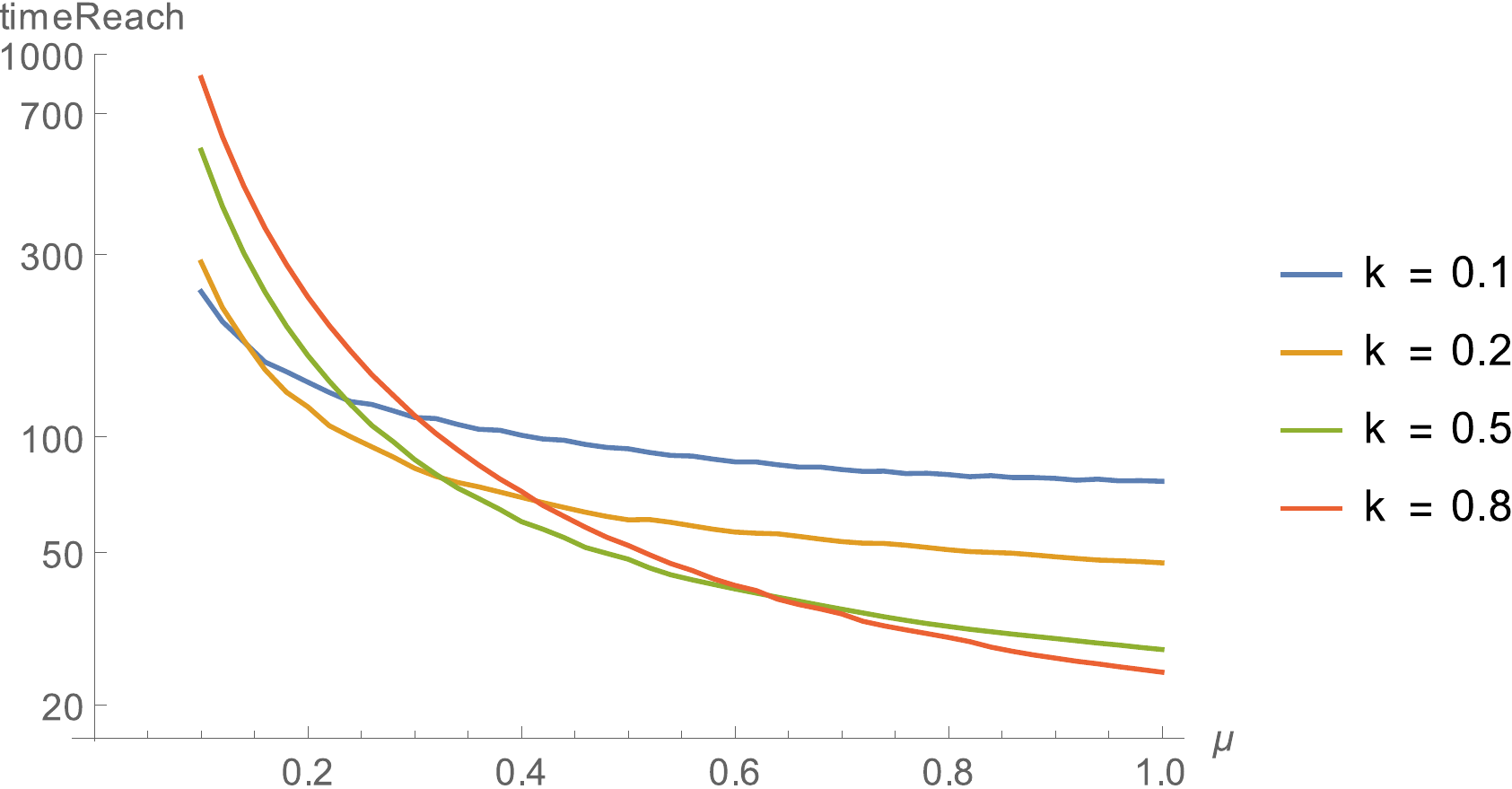} \\
The optimal output rate, over $\mu$ & & Time taken to reach target sink value $0.995$ \\
 & & using the optimal output rate, over $\mu$\\
\includegraphics[width=0.45\textwidth]{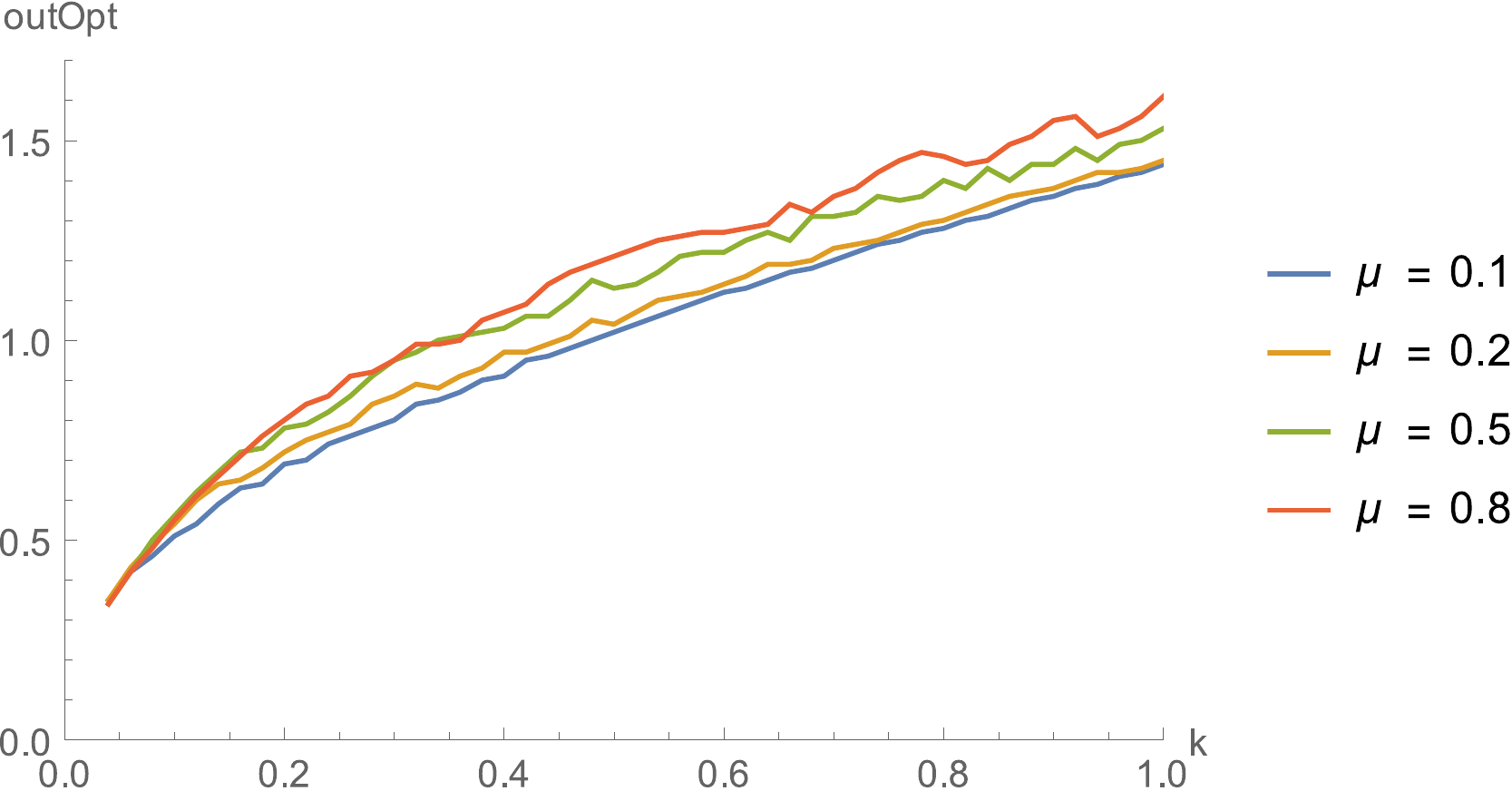} & & \includegraphics[width=0.45\textwidth]{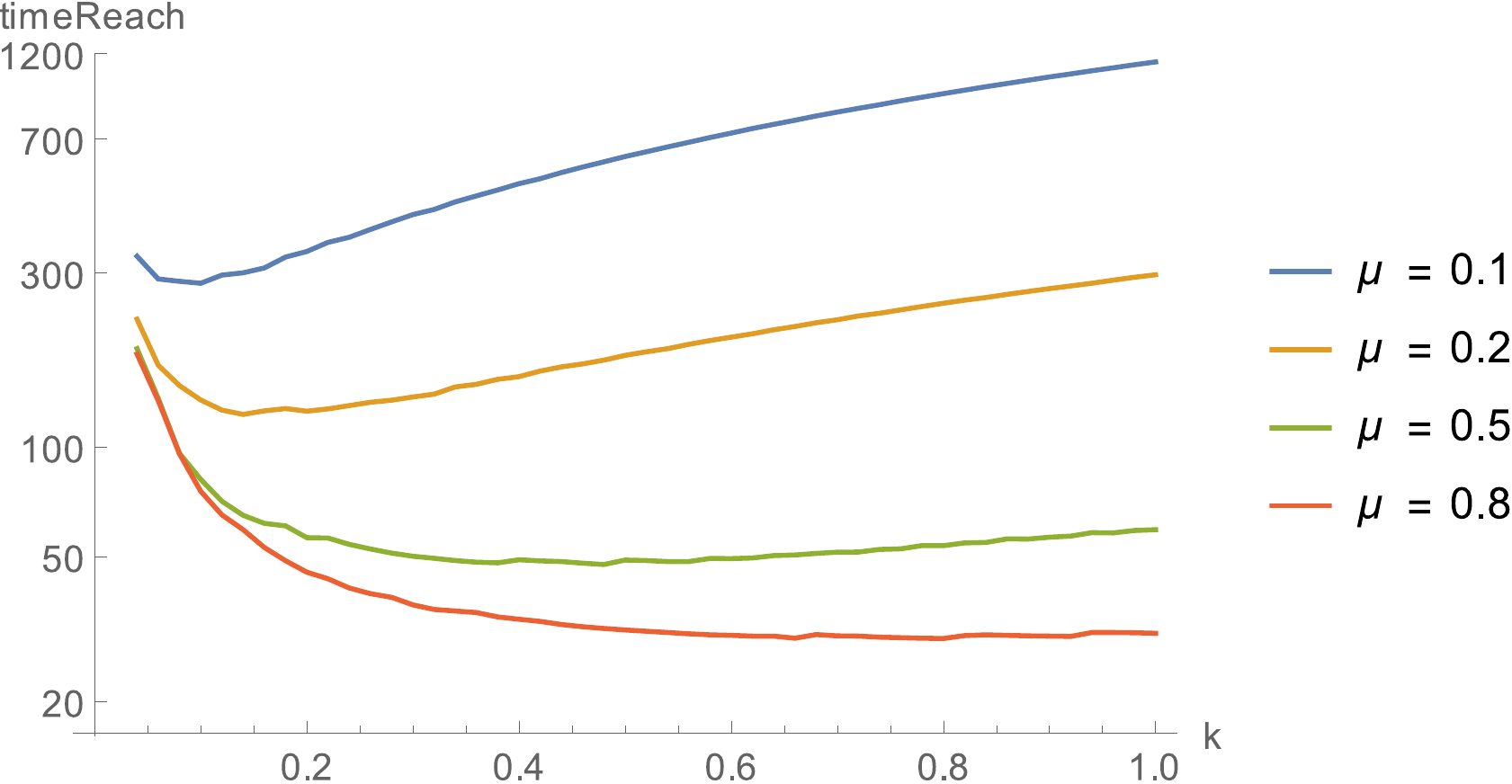} \\
The optimal output rate, over $k$ & & Time taken to reach target sink value $0.995$ \\
 & & using the optimal output rate, over $k$ \\
\end{tabular}
\vspace{0.2cm}
   \caption{ \label{fig:2-rkrfrfkr-noin} Optimal output rate for for a chain of two atoms without input (initial state has a photon in the first cavity). }
\end{figure}

\begin{figure}[H]
\begin{tabular}{c c c}
\includegraphics[width=0.45\textwidth]{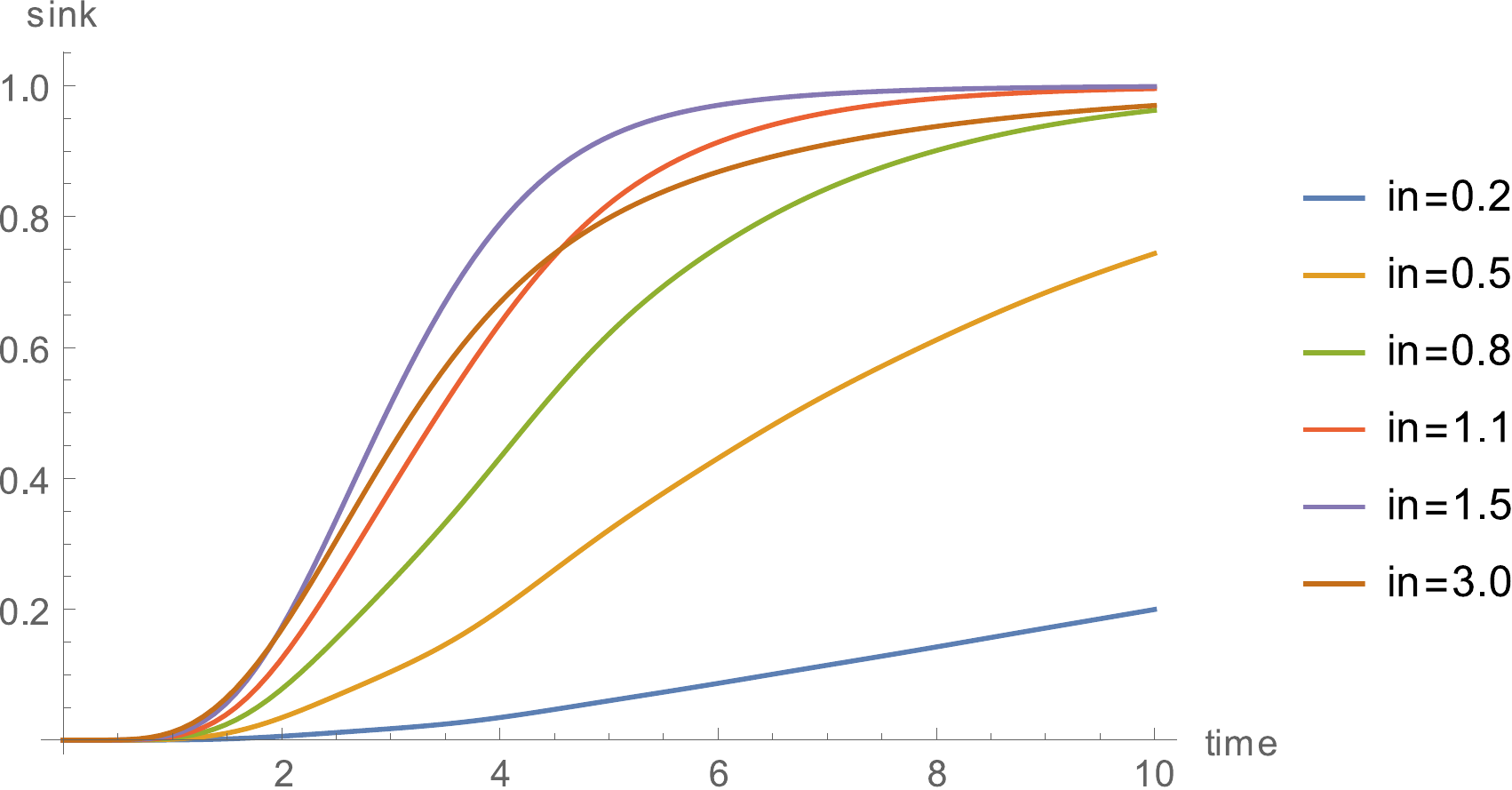} & & \includegraphics[width=0.45\textwidth]{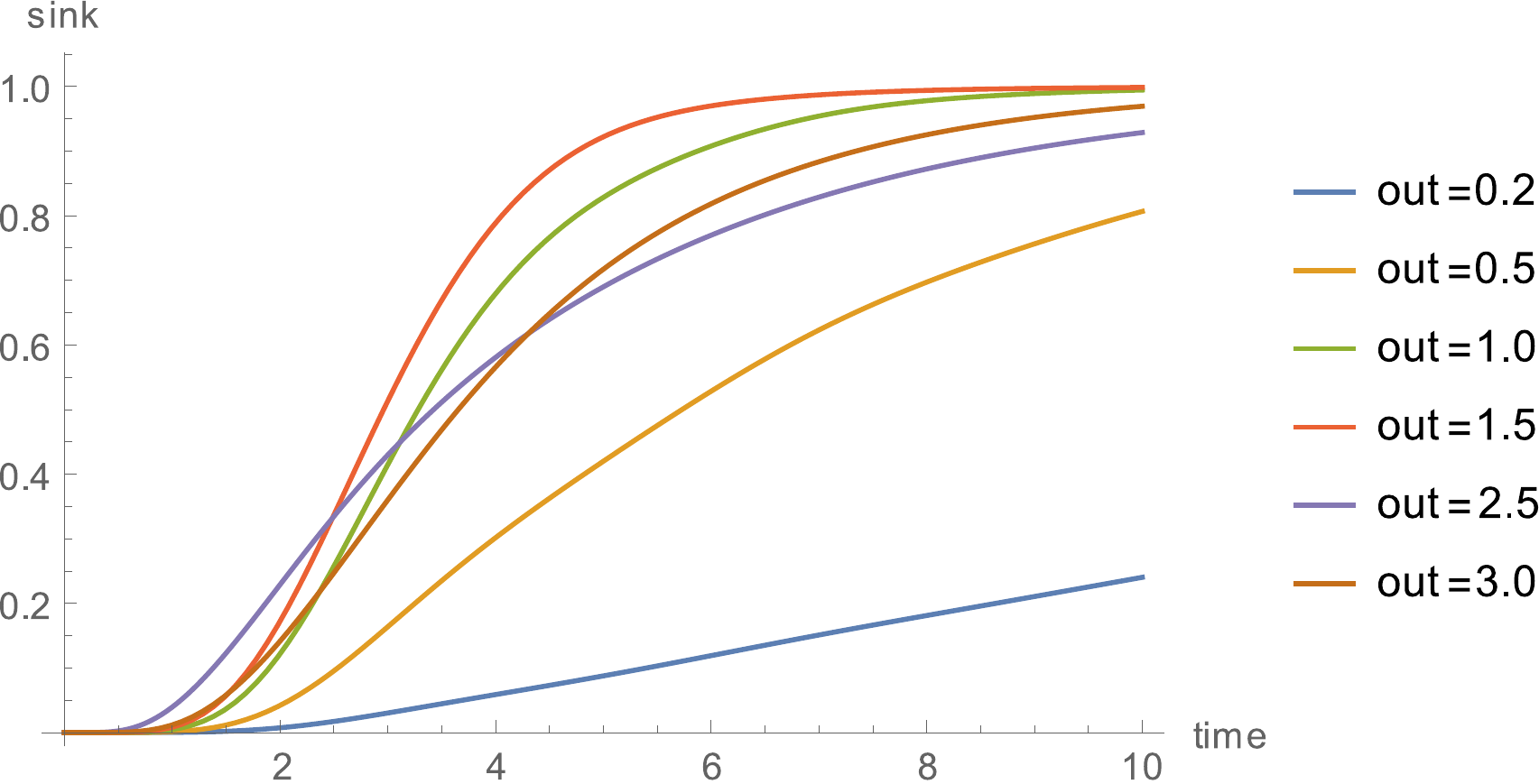} \\
$k$ = 1.0, $\mu$ = 1.0, $out$ = 1.5 (optimal) & & $k$ = 1.0, $\mu$ = 1.0, $in$ = 1.5 (optimal) \\
optimal input rate is 1.5 & & optimal output rate is 1.5 \\
\includegraphics[width=0.45\textwidth]{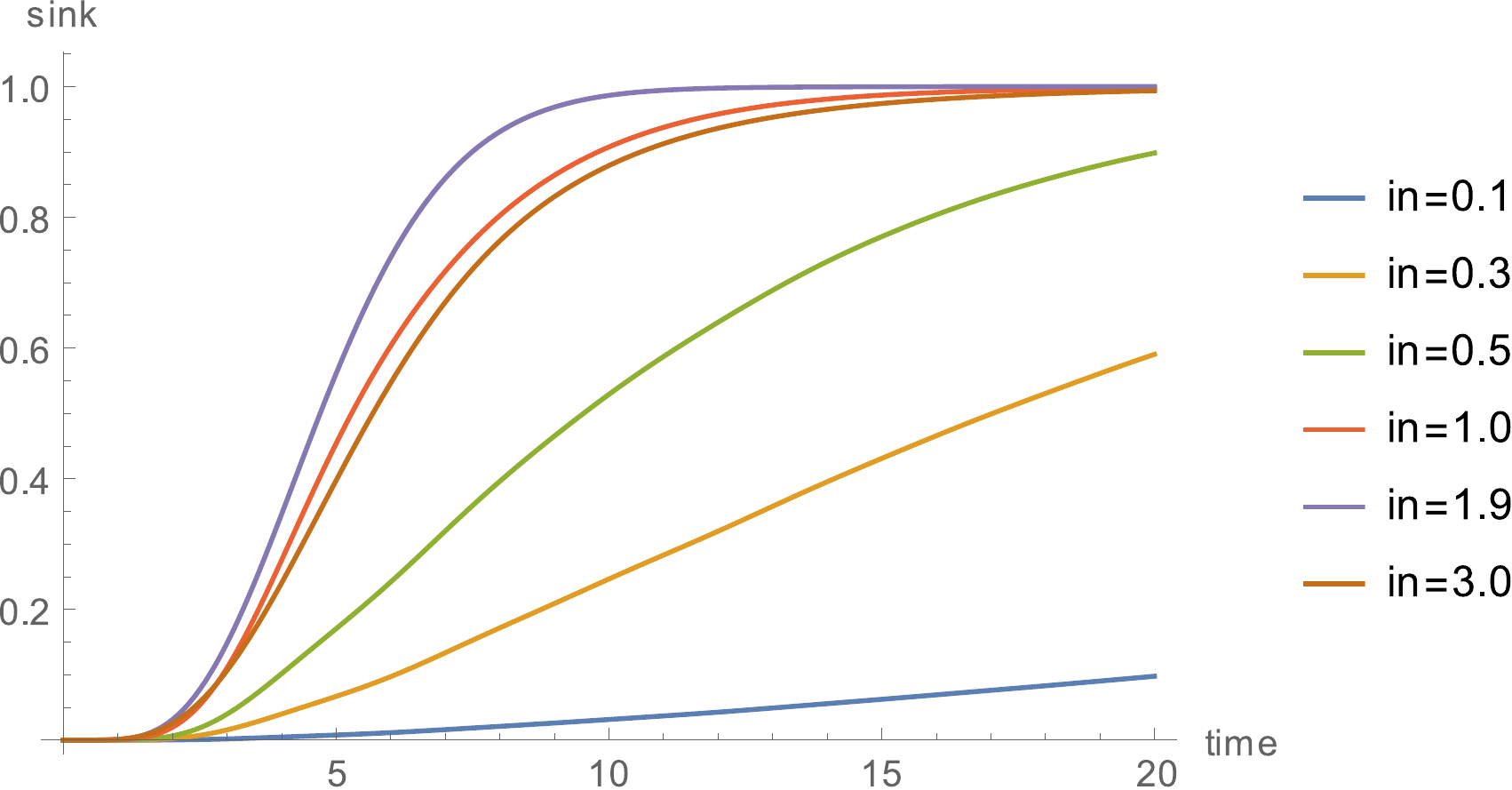} & & \includegraphics[width=0.45\textwidth]{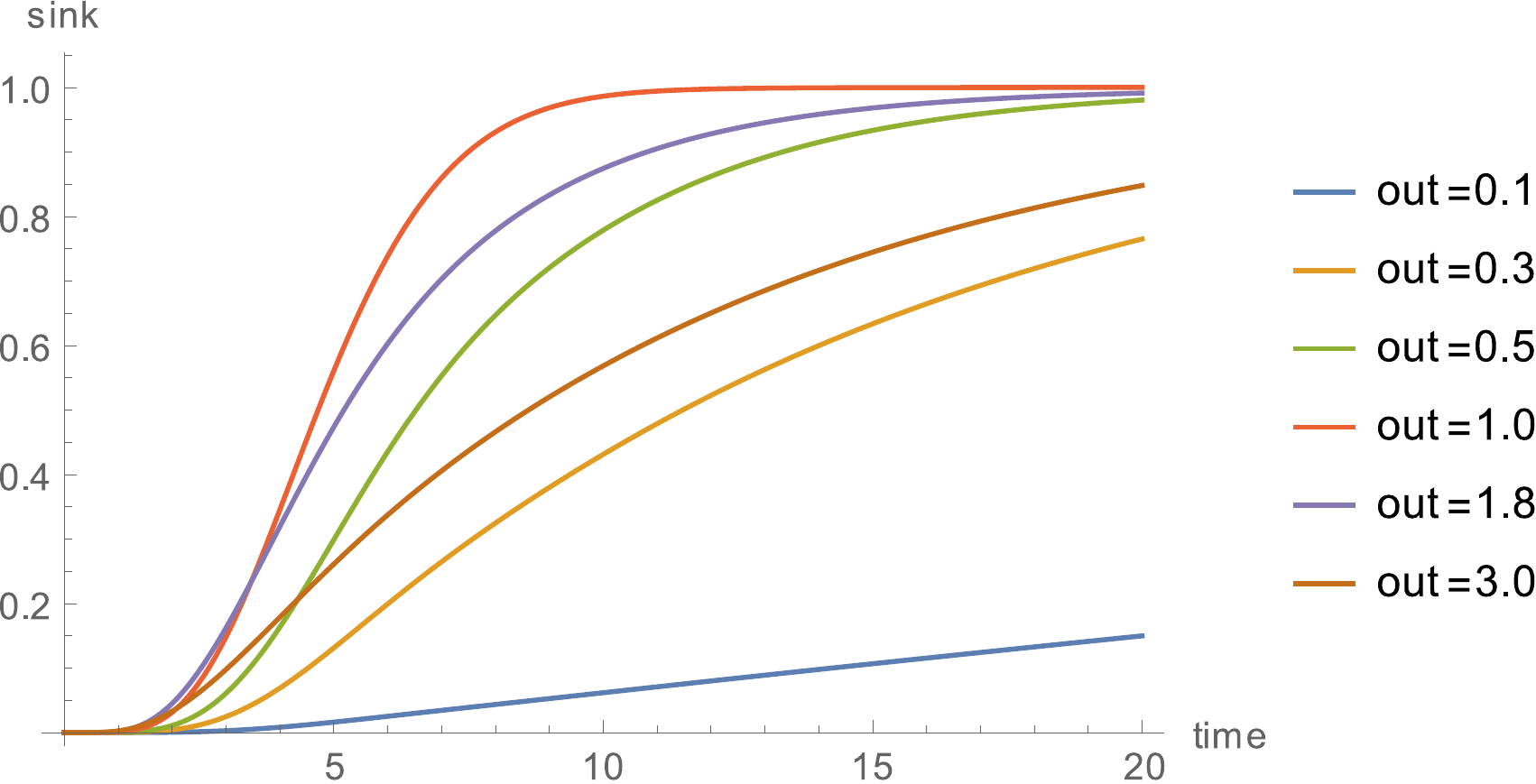} \\
$k$ = 0.8, $\mu$ = 0.5, $out$ = 1.0 (optimal) & & $k$ = 0.8, $\mu$ = 0.5, $in$ = 1.9 (optimal) \\
optimal input rate is 1.9 & & optimal output rate is 1.0 \\
\end{tabular}
\vspace{0.2cm}
   \caption{ \label{fig:2-bottleneck-time} Evolution of the sink state over time }
\end{figure}

Note that all two-dimensional $timeReach$ graphs in the paper have logarithmic scale.

For all numeric experiments with input enabled the total energy was not explicitly limited, except for the natural limits imposed by the model. For numeric experiments with input disabled, the initial state has one photon and the total energy is limited to $1$ (because it could not raise higher than that).

On figure \ref{fig:2-rkrfrfkr-noin} it is seen that for a chain of 2 atoms the optimal output rate has little dependency on the photon tunneling rate, but high dependency on photon-atom interaction strength.

It is also visible that increasing the photon-atom interaction strength over some critical value could have a negative effect on conductivity.

From the evolution graphs in figure \ref{fig:2-bottleneck-time} it could be seen that the optimal input/output rate depends on the target sink value (though that dependency is not critical of high enough target sink values). For example, with low target sink value $0.3$ and $k = 1.0, \mu = 1.0$, it is more effective to use $in = 1.5, out = 2.5$ than $in = 1.5, out = 1.5$.

\section{Dephasing-assisted transport}
\subsection{Lindbad-based dephasing model}
\label{lindblad-based-results}

\vspace{-0.1cm}
\begin{figure}[H]
\begin{tabular}{c c c}
\includegraphics[width=0.45\textwidth]{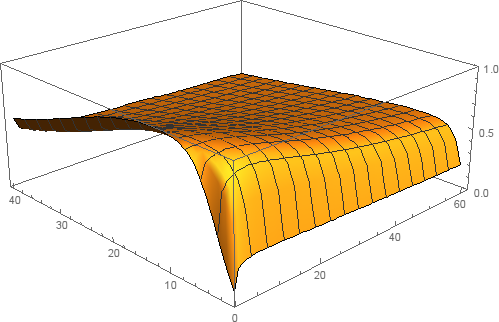} & & \includegraphics[width=0.45\textwidth]{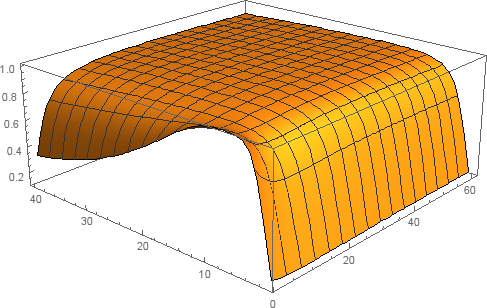} \\
$\mu$ = 0.2, $k$ = 0.8, $time$ = 150 & &  $\mu$ = 0.8, $k$ = 0.2, $time$ = 60 \\
\end{tabular}
\vspace{0.0cm}
\caption{ \label{fig:2-dat-inout-3d} The state of sink at the fixed time. Two atoms, no input (initial state has a photon in the first cavity). Dependency over the output rate $out$ and the dephasing coefficient $g$.}
\end{figure}

\vspace{-0.4cm}

For high enough times, the absolute maximum over all possible output rates is reached when $g = 0$. But non-zero values of $g$ can greatly improve the conductivity in cases when the output rate does not match with the optimal output rate. That is shown on figure \ref{fig:2-dat-inout-3d}. It could also be seen that the effect does not depend on $g$ being equal to some exact value in some cases, and there is a broad range of possible values for $g$ that improves the conductivity.

For very low times (or low target sink values, depending on the stop criterion) these graphs are different and could show other short-lived effects, but we are inspecting target sink values close to $1.0$.

In the first part of \ref{fig:2-dat-inout-3d} ($\mu$ = 0.2, $k$ = 0.8) it could be seen that non-zero values for $g$ improve the conductivity for the case $out < outOpt$. In the second part ($\mu$ = 0.8, $k$ = 0.2) it could be seen that non-zero values for $g$ improve the conductivity for the case $out > outOpt$.

The exact range of possible output rate values for which the conductivity could be improved by adding dephasing depends on the parameters of the chain.

\begin{figure}[H]
\begin{tabular}{c c c}
\includegraphics[width=0.45\textwidth]{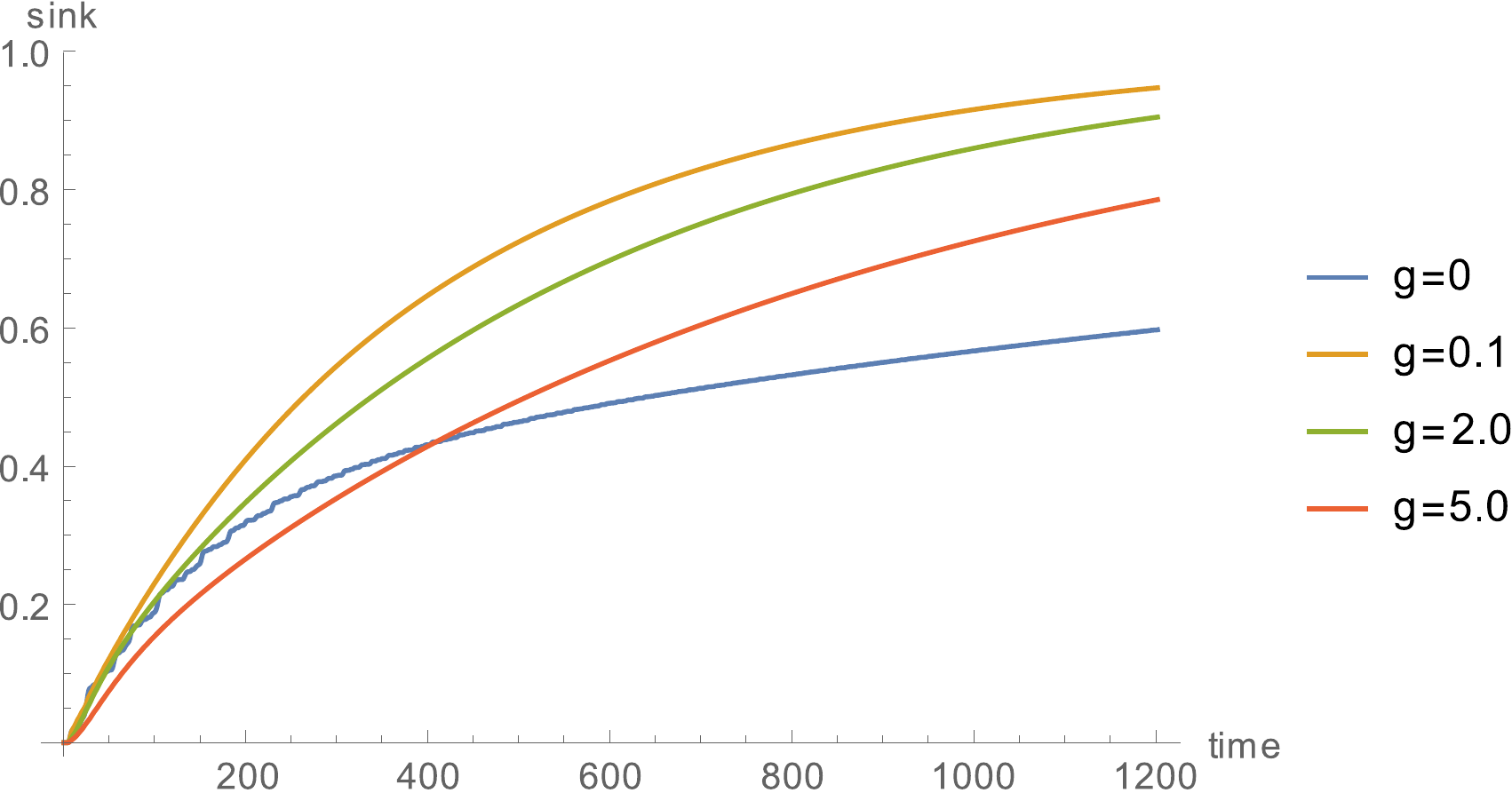} & & \includegraphics[width=0.45\textwidth]{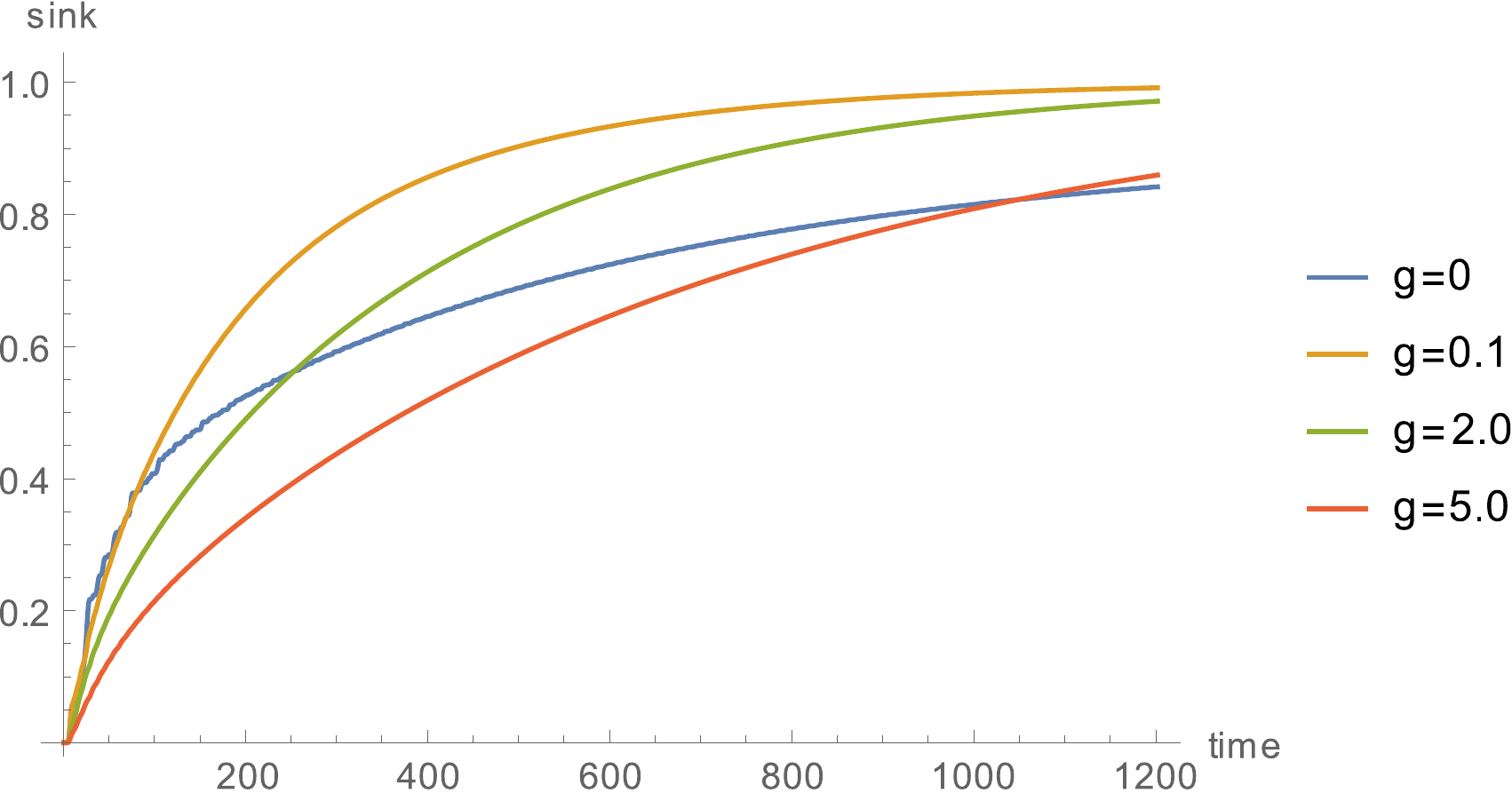} \\
$out$ = 0.2 & & $out$ = 0.4 \\
\includegraphics[width=0.45\textwidth]{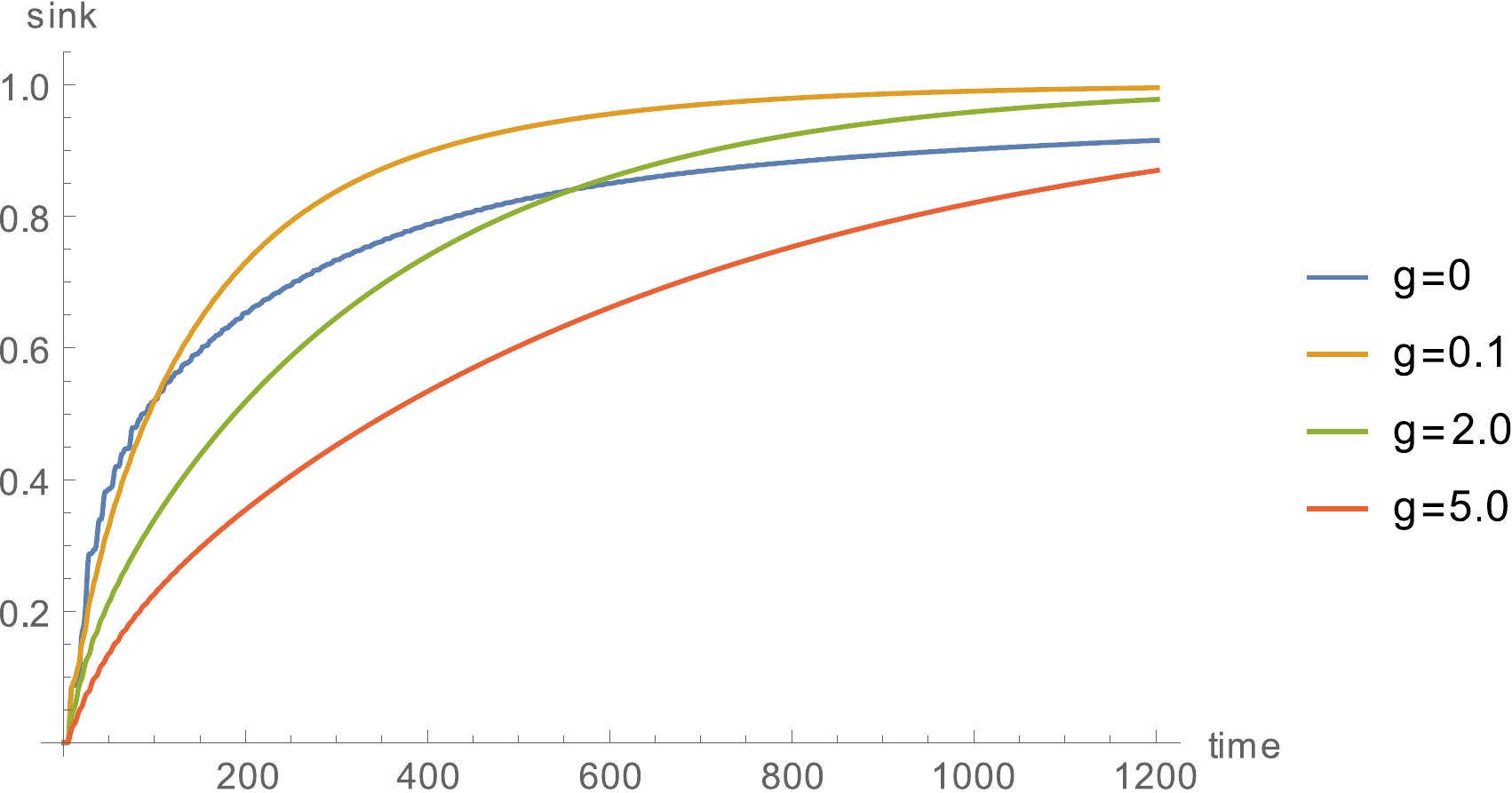} & & \includegraphics[width=0.45\textwidth]{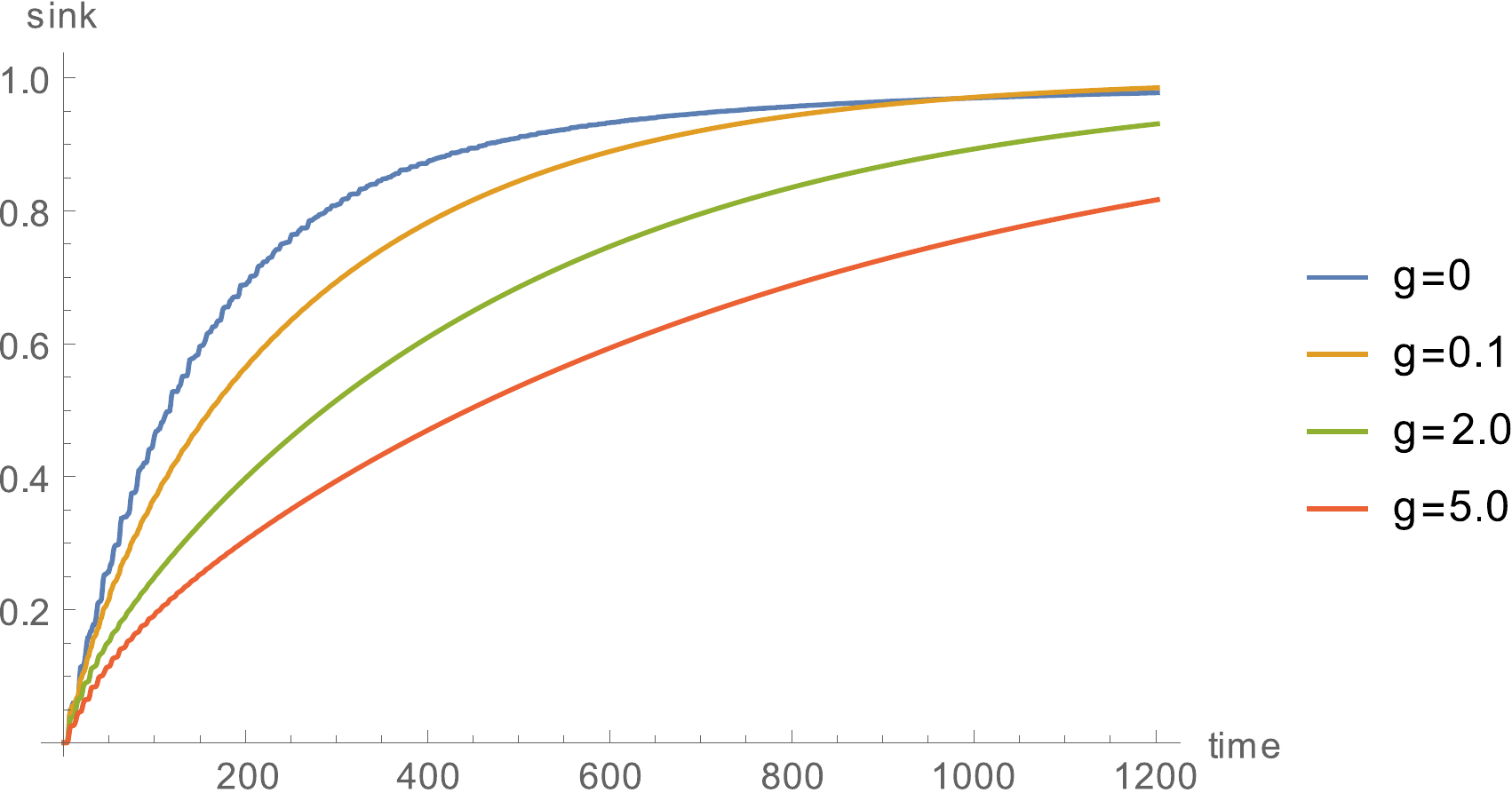} \\
$out$ = 0.6 & & $out$ = 2.0 \\
\end{tabular}
\vspace{0.2cm}
 \caption{ \label{fig:2-dat-time2} Evolution of the sink state over time. Five atoms, no input (initial state has a photon in the first cavity). $k$ = 0.8, $\mu$ = 0.2. }
\end{figure}

The same effects could be observed for higher chain lengths, as could be seen on figure \ref{fig:2-dat-time2}.

\begin{figure}[H]
\begin{tabular}{c c c}
\includegraphics[width=0.45\textwidth]{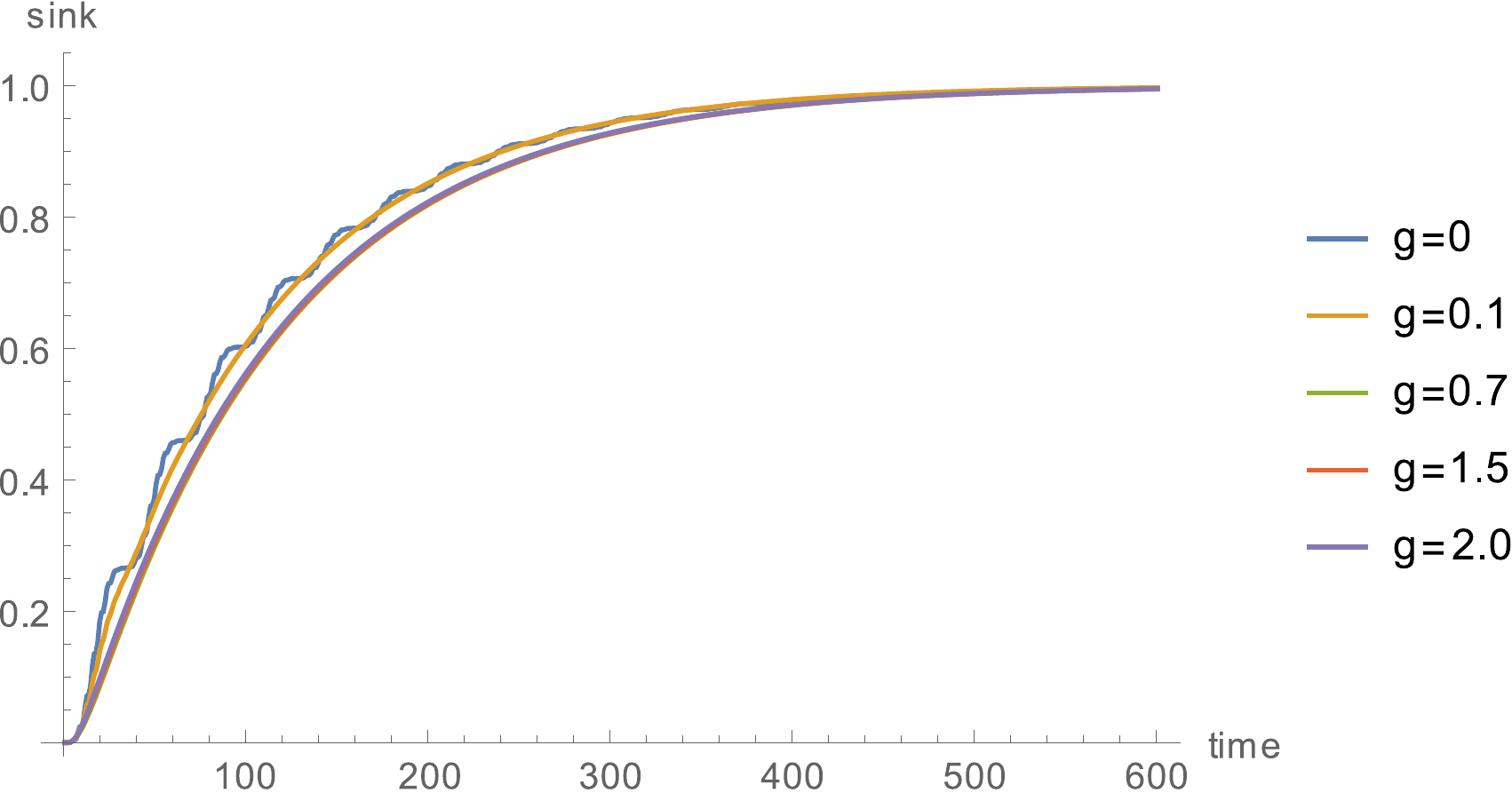} & & \includegraphics[width=0.45\textwidth]{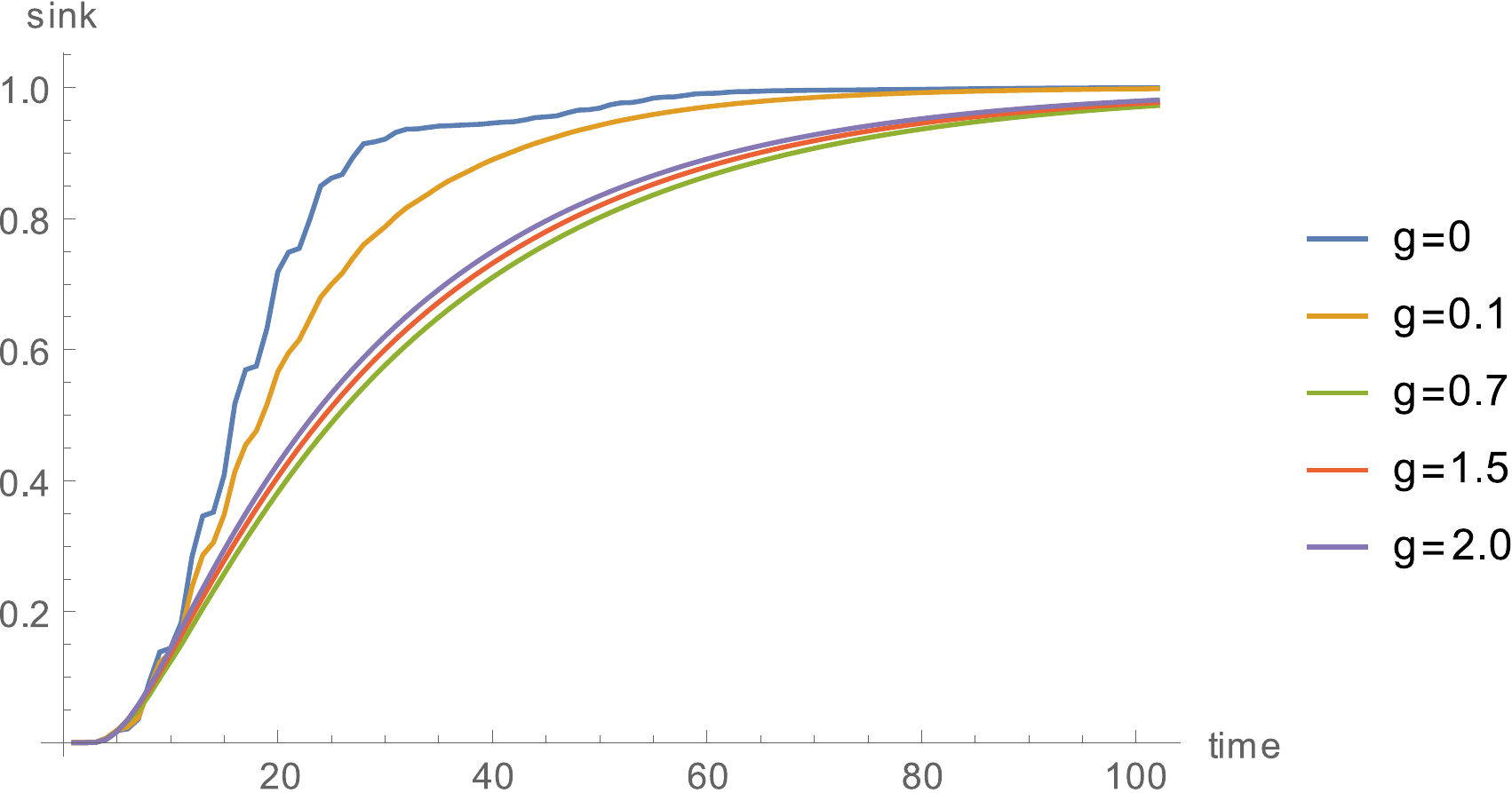} \\
$out$ = 0.2 & & $out$ = 0.6 \\
\includegraphics[width=0.45\textwidth]{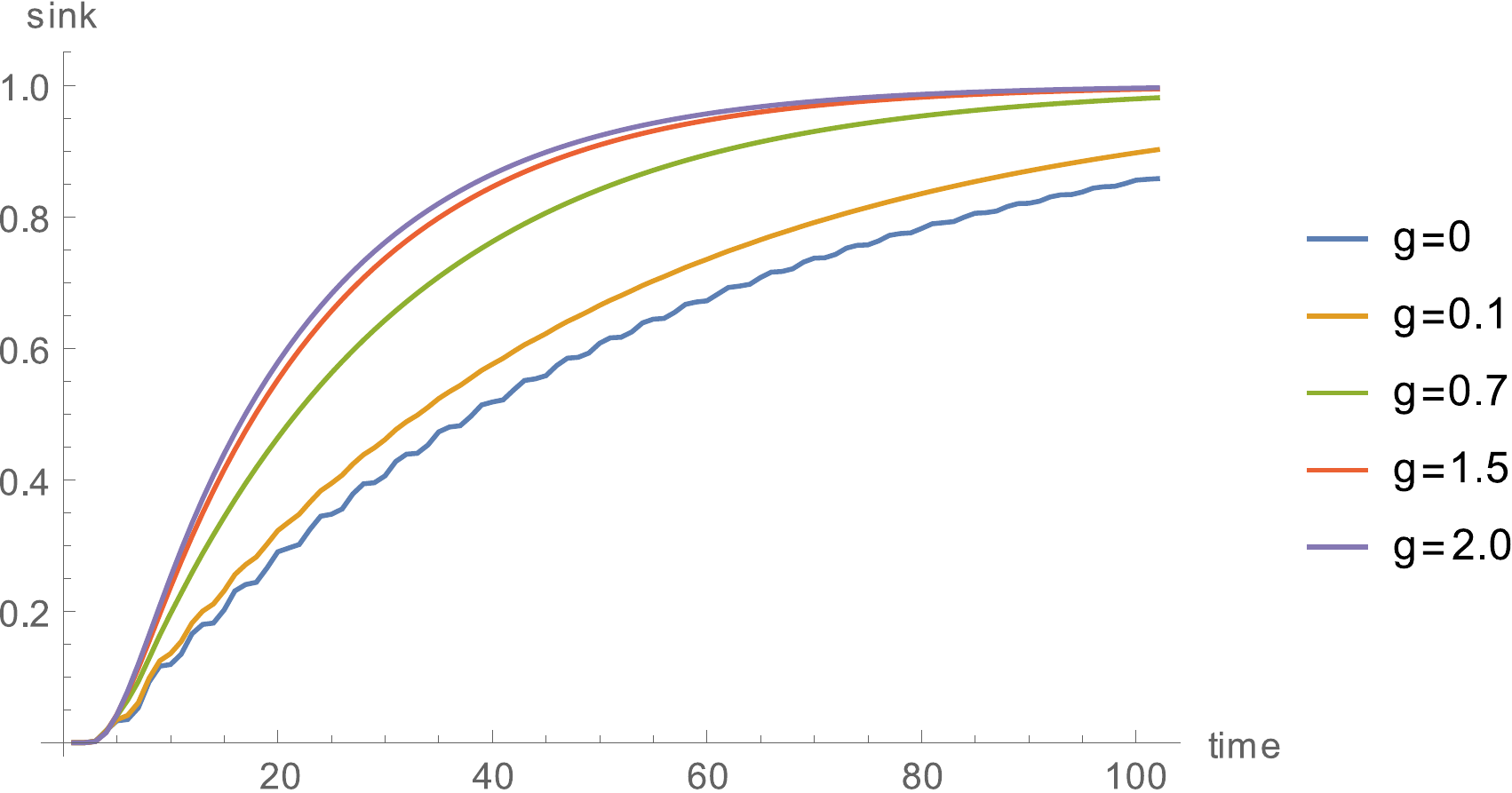} & & \includegraphics[width=0.45\textwidth]{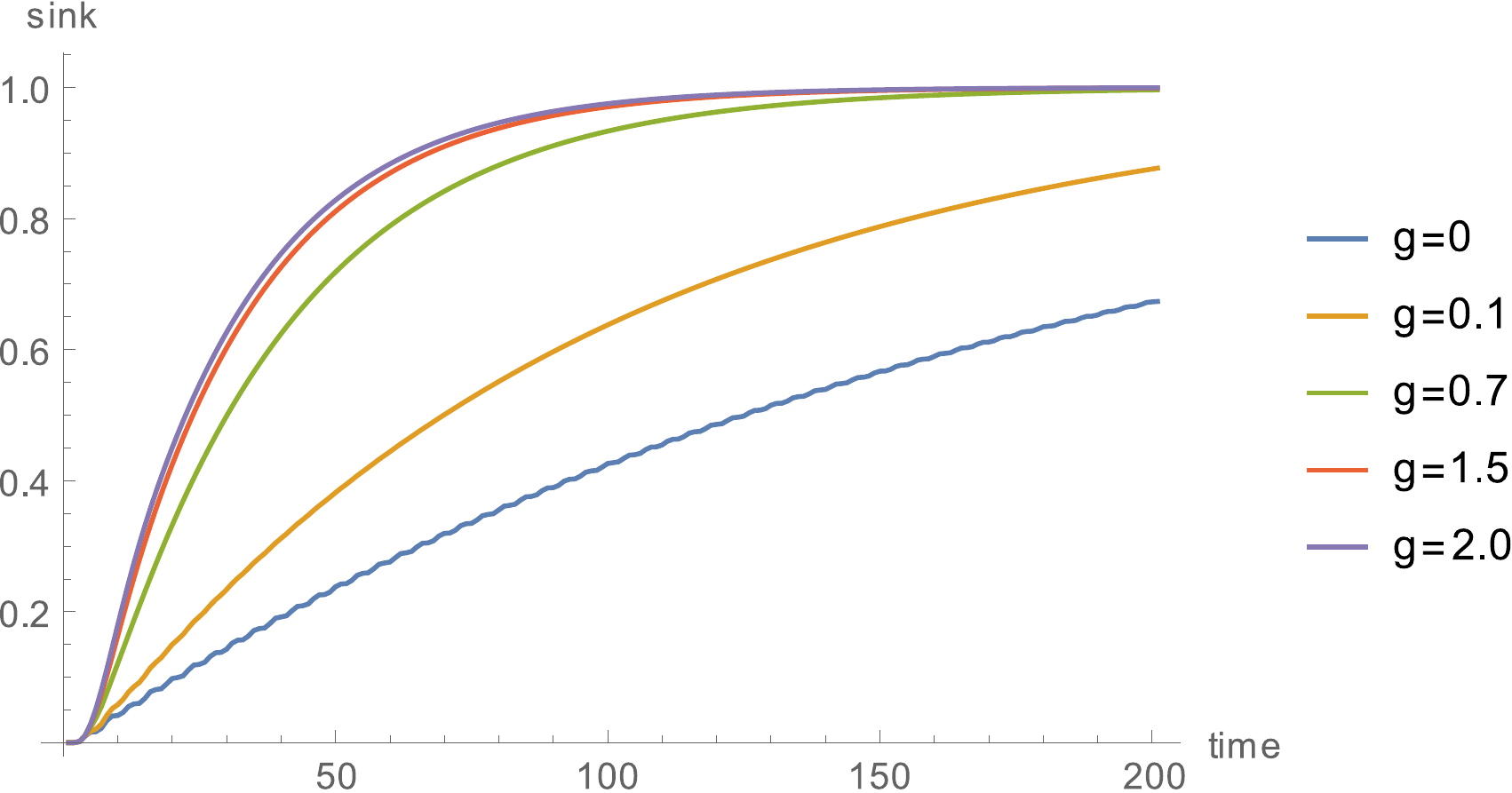} \\
$out$ = 2.0 & & $out$ = 4.0 \\
\end{tabular}
\vspace{0.2cm}
 \caption{ \label{fig:2-dat-time0} Evolution of the sink state over time. Two atoms, no input (initial state has a photon in the first cavity). $k$ = 0.2, $\mu$ = 0.8. }
\end{figure}

\begin{figure}[H]
\begin{tabular}{c c c}
\includegraphics[width=0.45\textwidth]{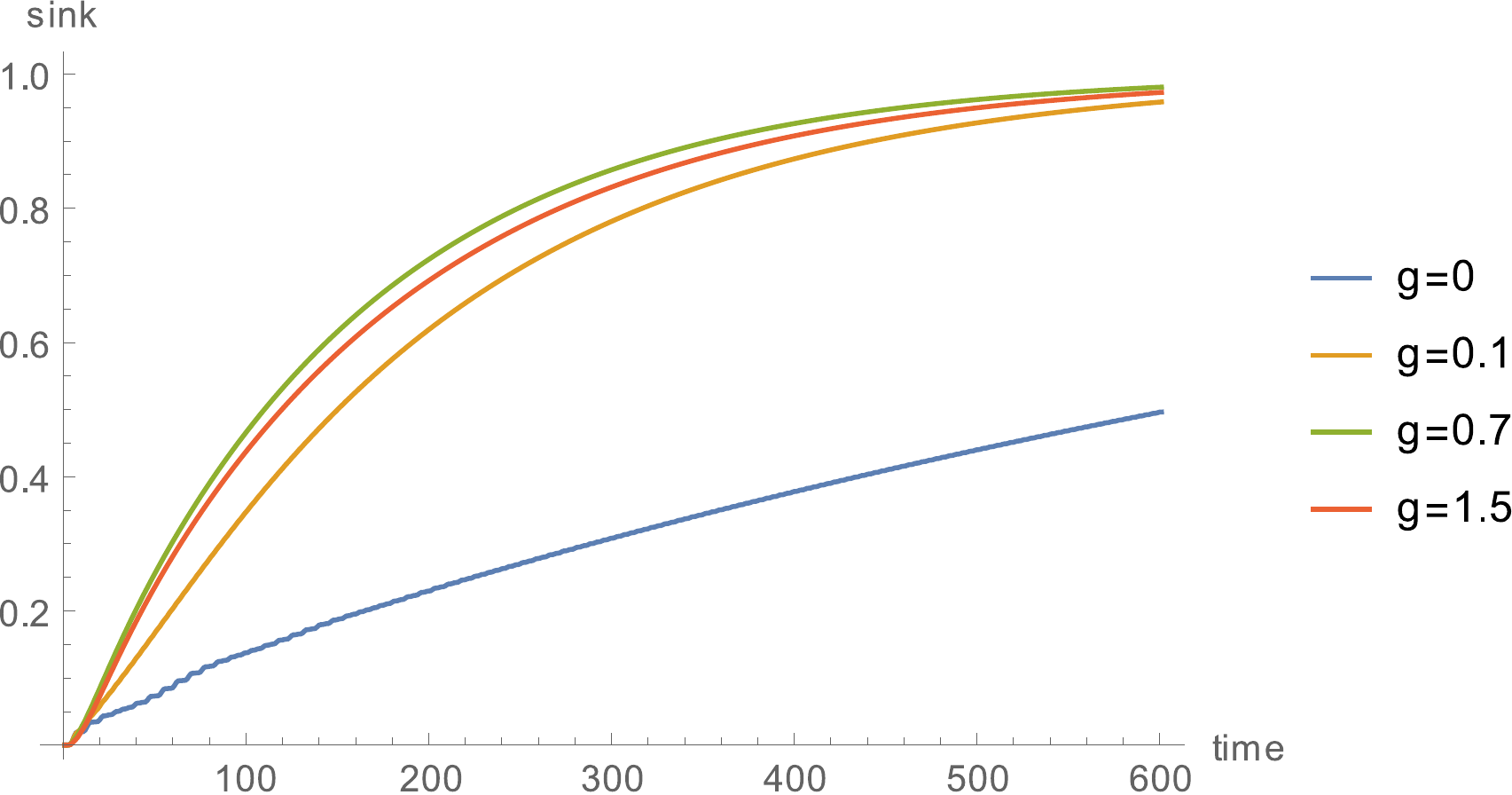} & & \includegraphics[width=0.45\textwidth]{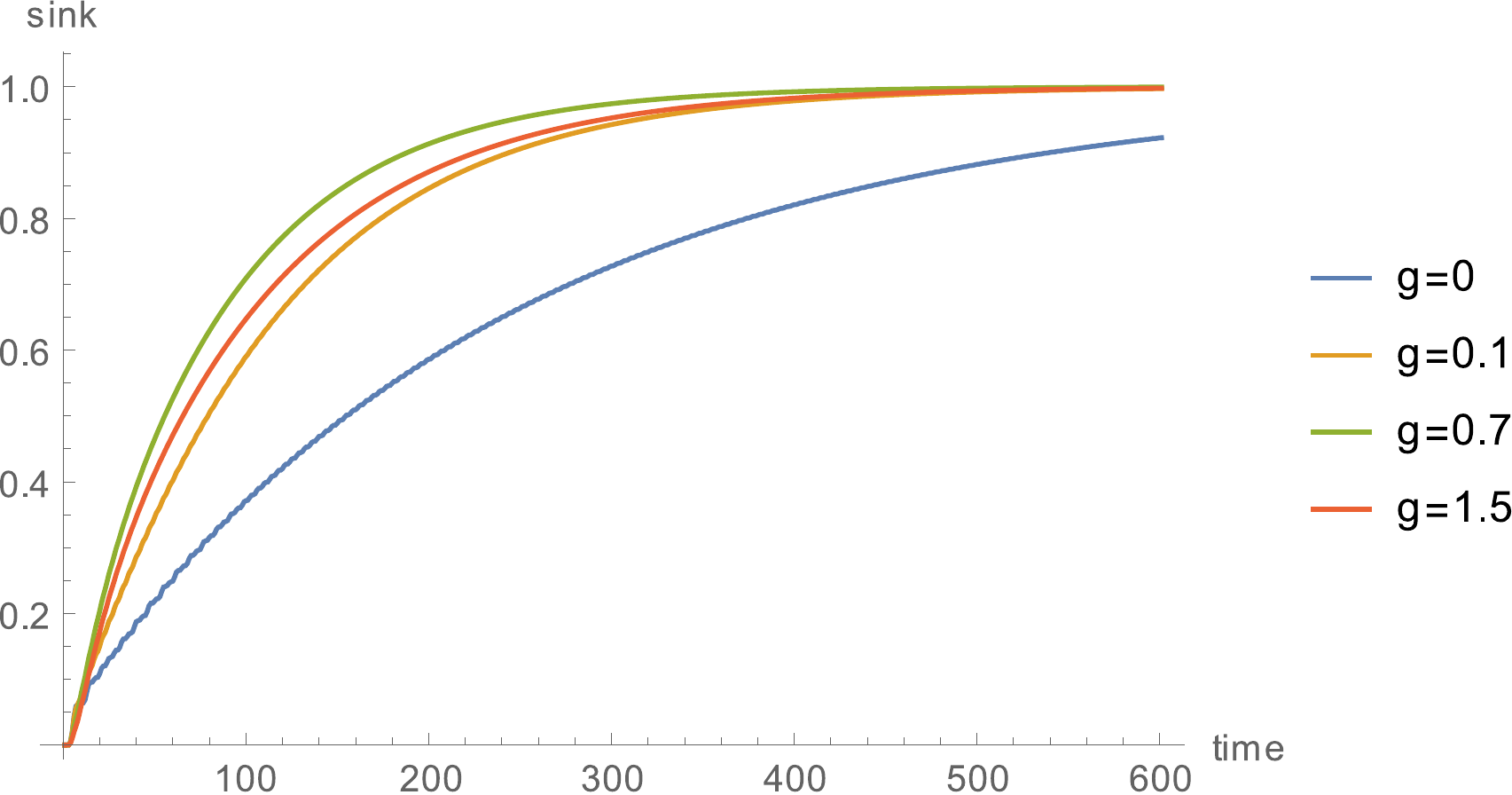} \\
$out$ = 0.2 & & $out$ = 0.4 \\
\includegraphics[width=0.45\textwidth]{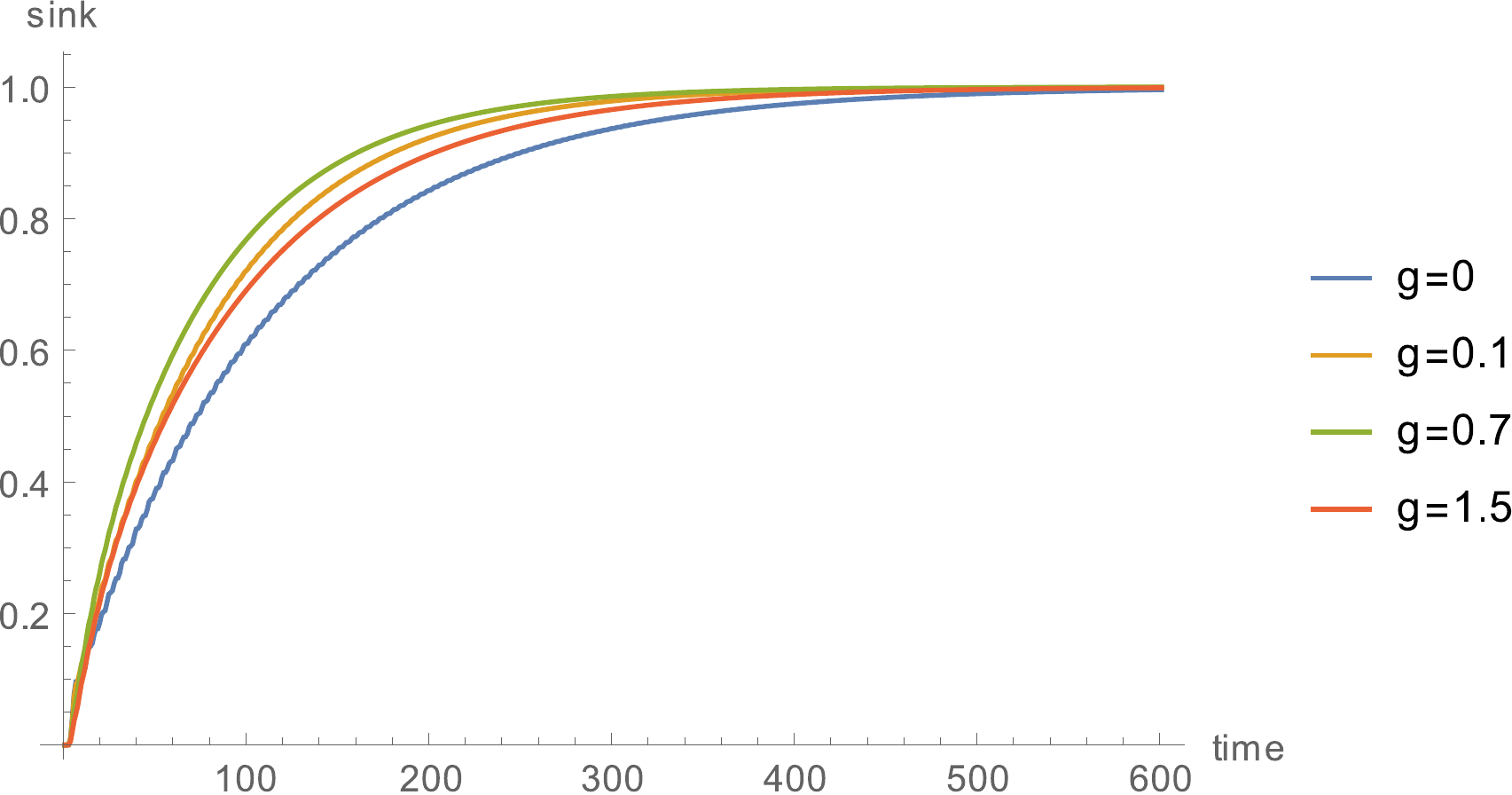} & & \includegraphics[width=0.45\textwidth]{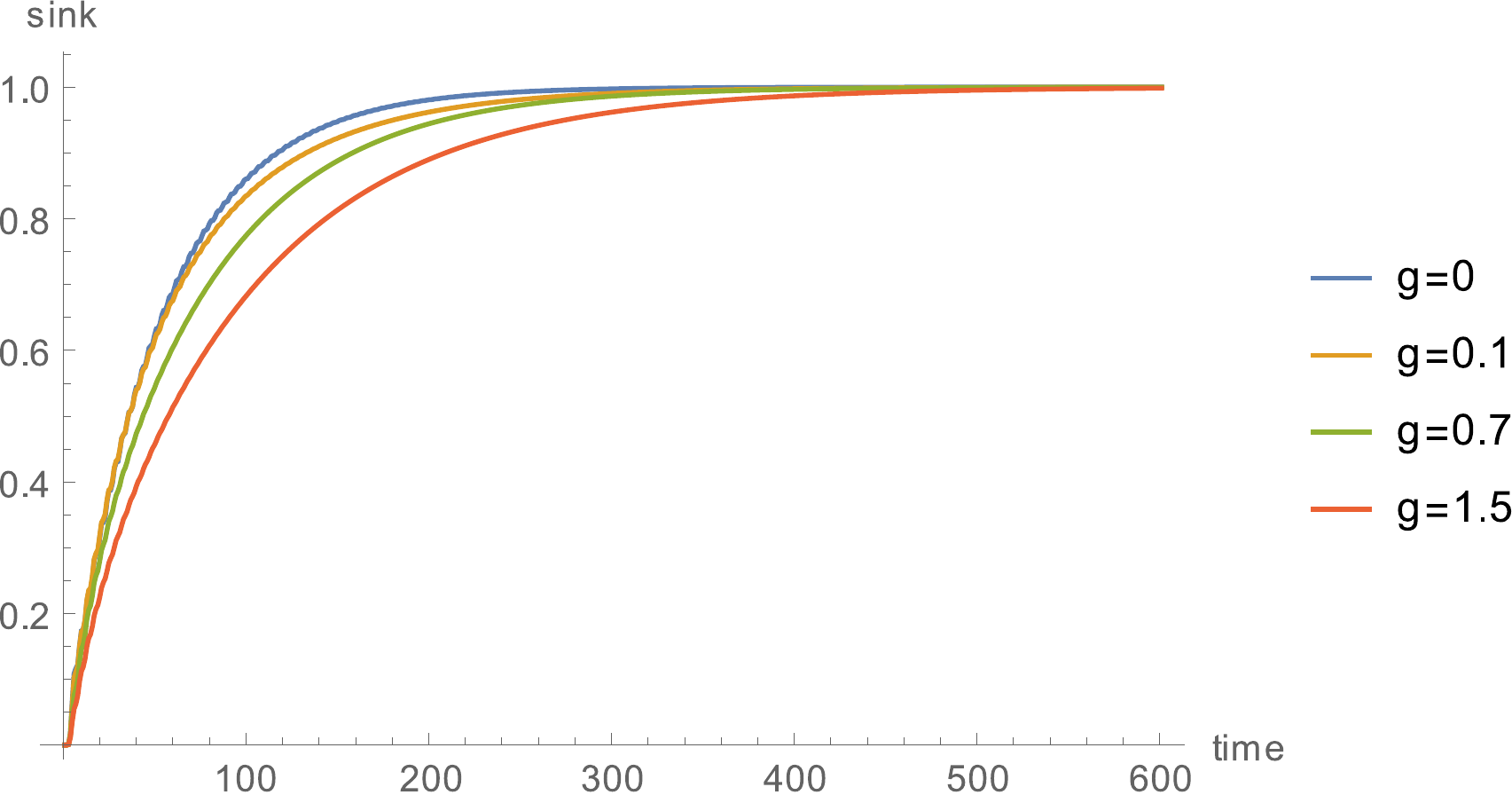} \\
$out$ = 0.6 & & $out$ = 1.0 \\
\end{tabular}
\vspace{0.2cm}
 \caption{ \label{fig:2-dat-time1} Evolution of the sink state over time. Two atoms, no input (initial state has a photon in the first cavity). $k$ = 0.8, $\mu$ = 0.2. }
\end{figure}

\vspace{-0.3cm}

\subsection{Unitary dephasing model}

\begin{figure}[H]
\begin{tabular}{c c c}
\includegraphics[width=0.45\textwidth]{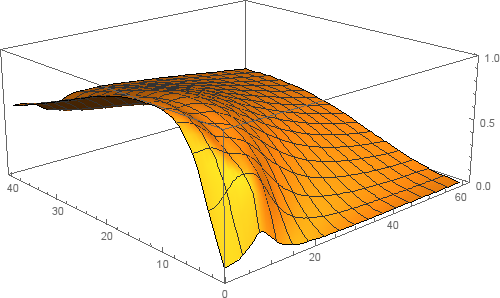} & & \includegraphics[width=0.45\textwidth]{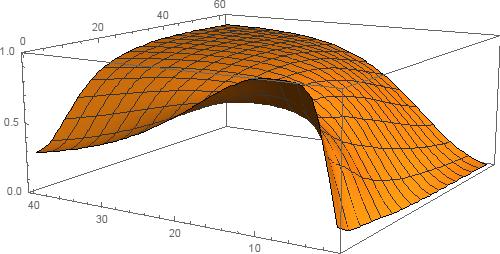} \\
$\mu$ = 0.2, $k$ = 0.8, $time$ = 150 & &  $\mu$ = 0.8, $k$ = 0.2, $time$ = 60 \\
\end{tabular}
\vspace{0.2cm}
\caption{ \label{fig:2-udat-inout-3d} The state of sink at the fixed time. Two atoms, no input (initial state has a photon in the first cavity). Dependency over the output rate $out$ and the dephasing coefficient $g$.}
\end{figure}

For the unitary dephasing model, the main observed effects are the same as for the Lindblad-based dephasing model in \ref{lindblad-based-results} --- non-zero values of $g$ can improve the conductivity in cases of inoptimal values of the output rate, though the overall effect of unitary dephasing (with the supplied parameters) is milder.

Figure \ref{fig:2-udat-inout-3d} depicts the same experiment as figure \ref{fig:2-dat-inout-3d}, but with the unitary dephasing model.

\begin{figure}[H]
\begin{tabular}{c c c}
\includegraphics[width=0.45\textwidth]{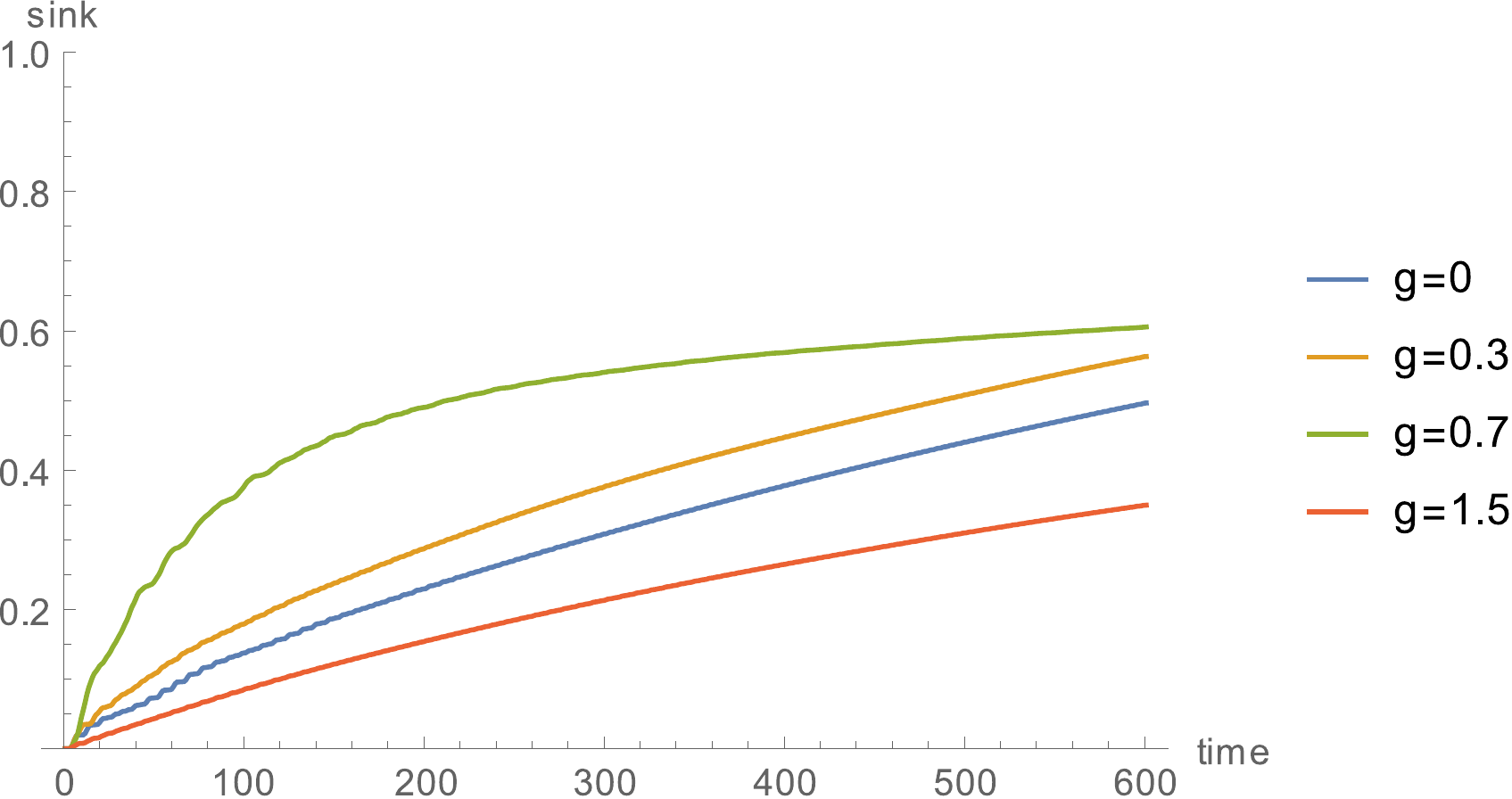} & & \includegraphics[width=0.45\textwidth]{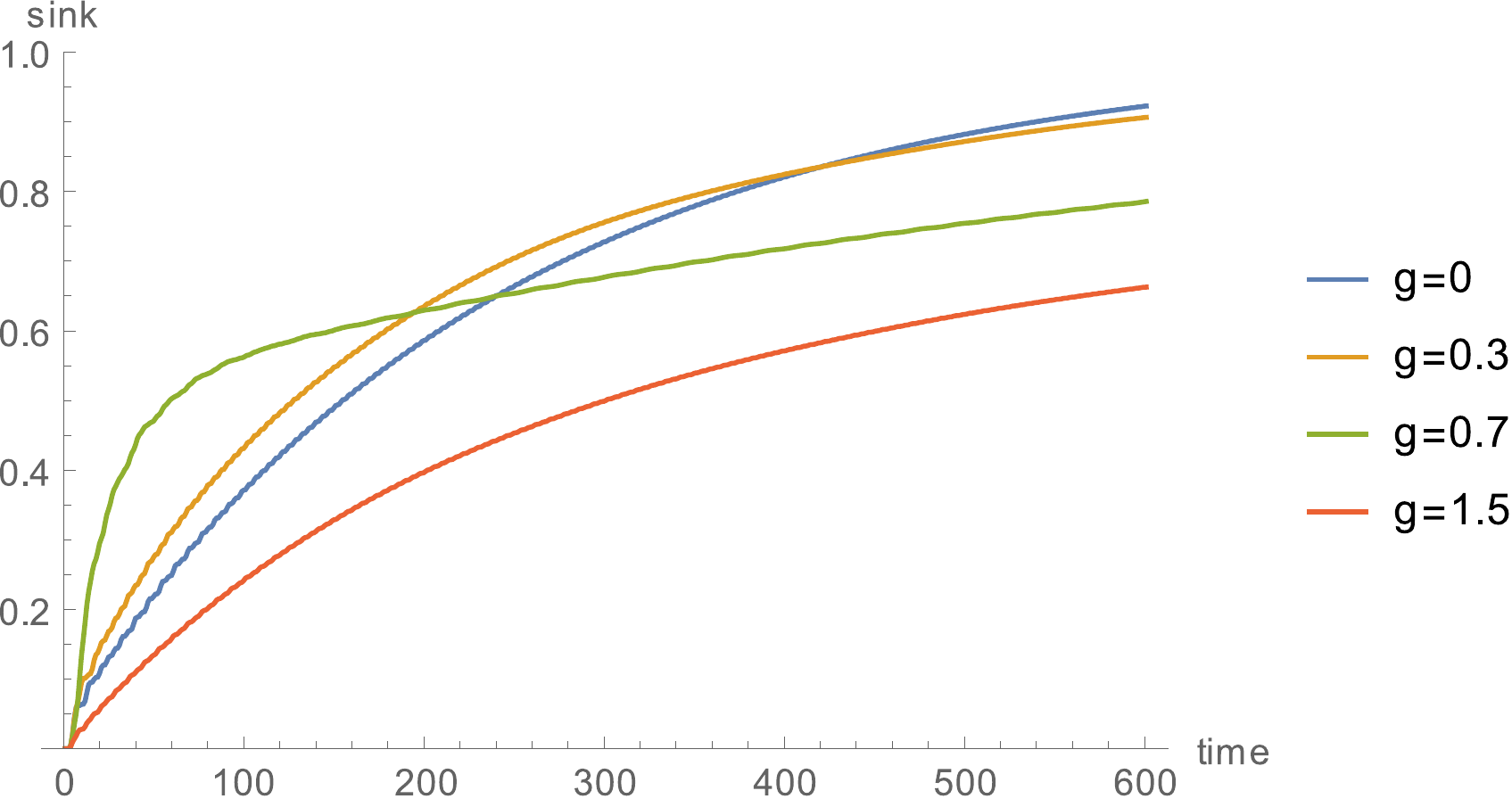} \\
$out$ = 0.2 & & $out$ = 0.4 \\
\includegraphics[width=0.45\textwidth]{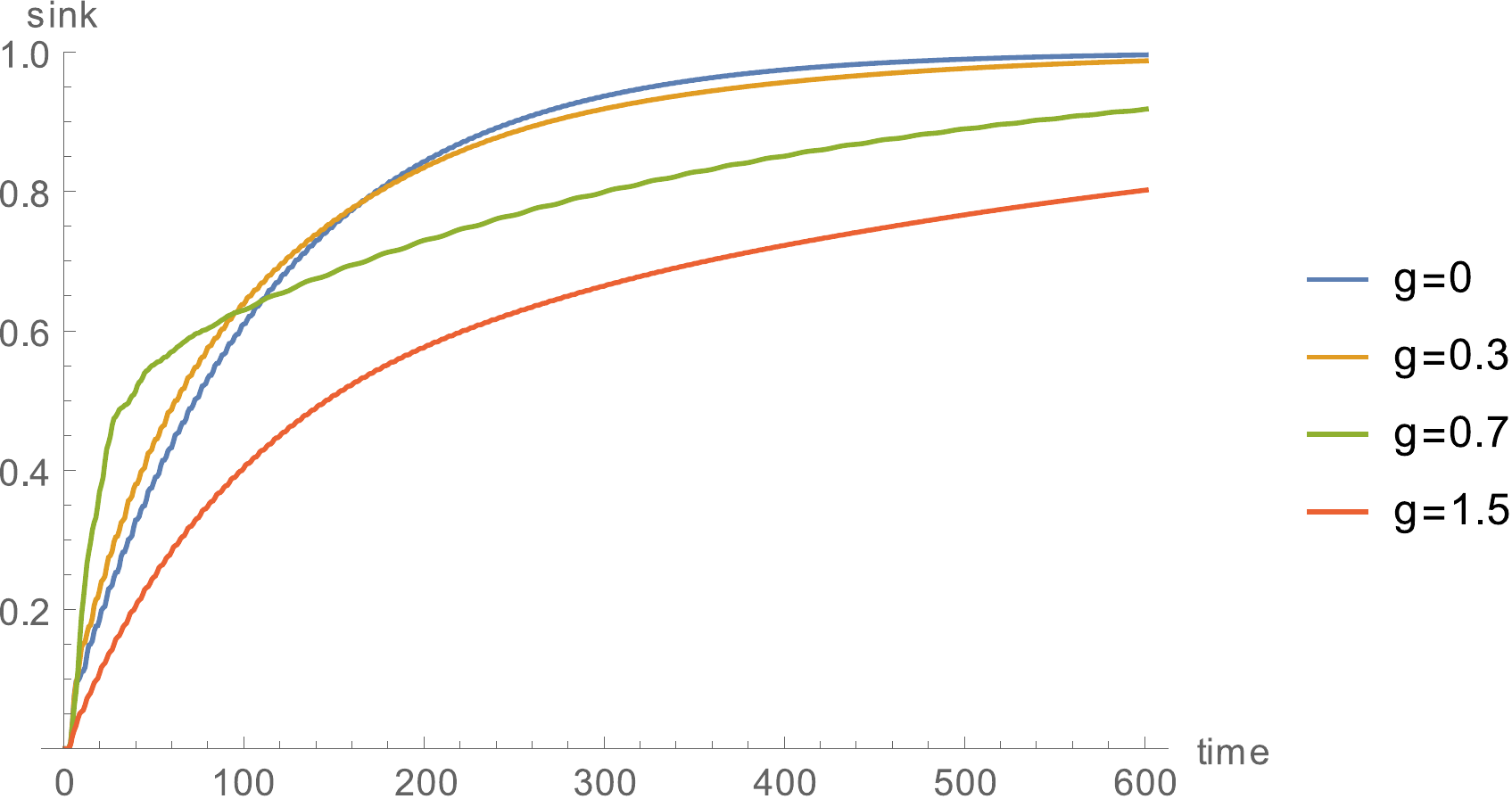} & & \includegraphics[width=0.45\textwidth]{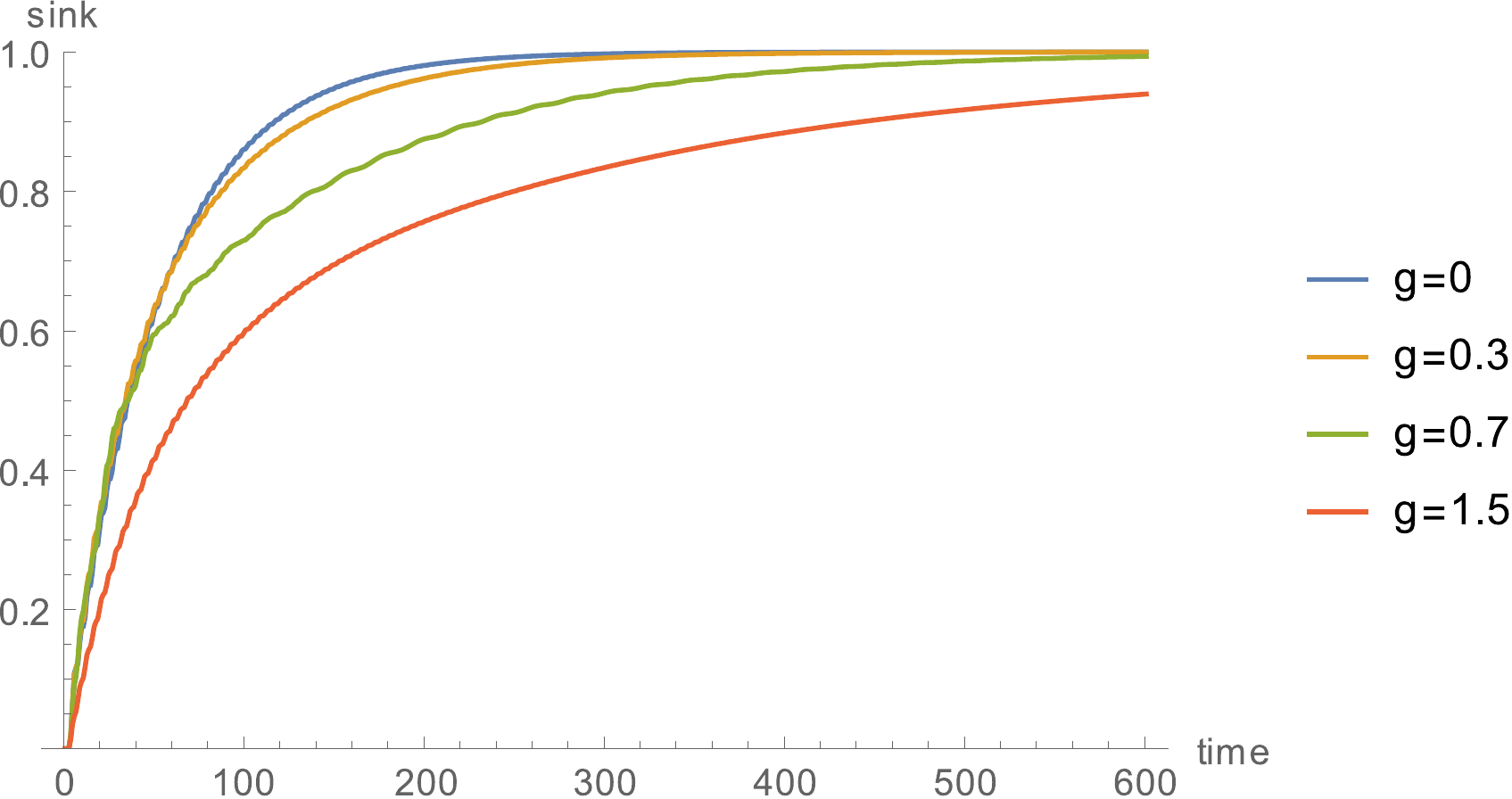} \\
$out$ = 0.6 & & $out$ = 1.0 \\
\end{tabular}
\vspace{0.2cm}
 \caption{ \label{fig:2-udat-time1} Evolution of the sink state over time. Two atoms, no input (initial state has a photon in the first cavity). $k$ = 0.8, $\mu$ = 0.2. }
\end{figure}

On figures \ref{fig:2-bottleneck-time}, \ref{fig:2-dat-time1}, \ref{fig:2-udat-time1} and it could be seen that setups that were optimal to reach low target sink values (for example $0.3$ or $0.4$) are not always optimal to reach target sink values close to $1$.

\begin{figure}[H]
\begin{tabular}{c c c}
\includegraphics[width=0.45\textwidth]{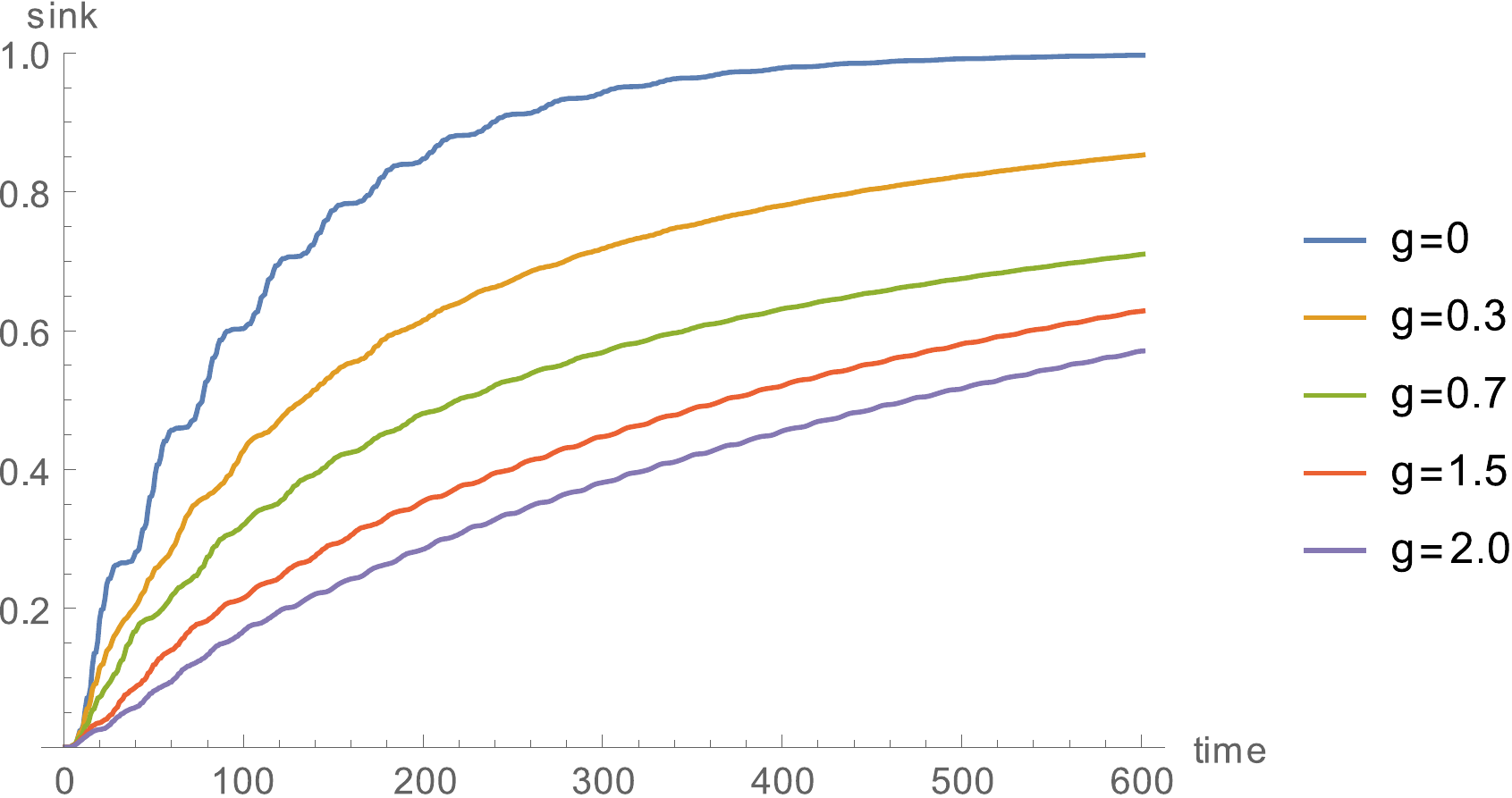} & & \includegraphics[width=0.45\textwidth]{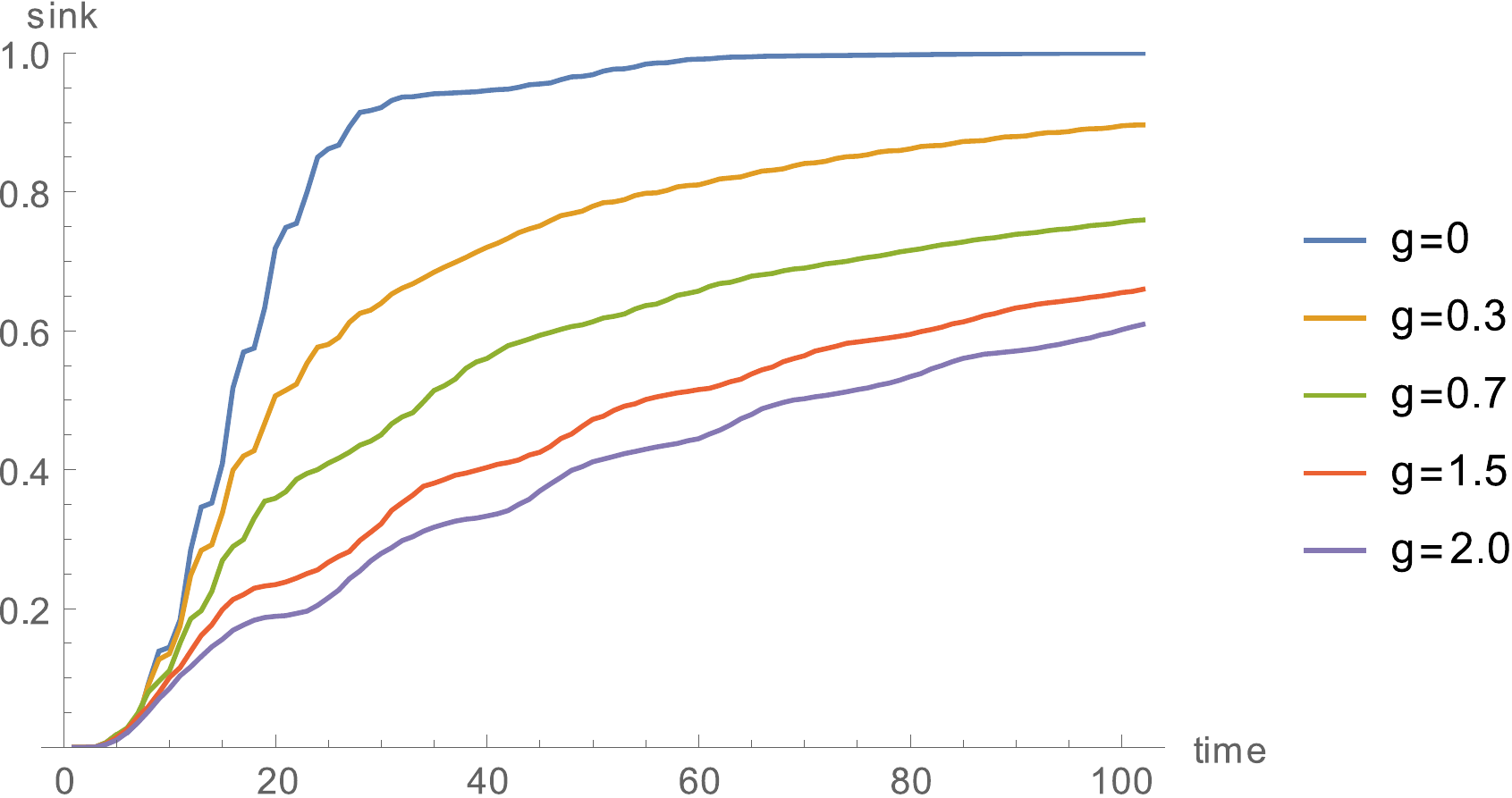} \\
$out$ = 0.2 & & $out$ = 0.6 \\
\includegraphics[width=0.45\textwidth]{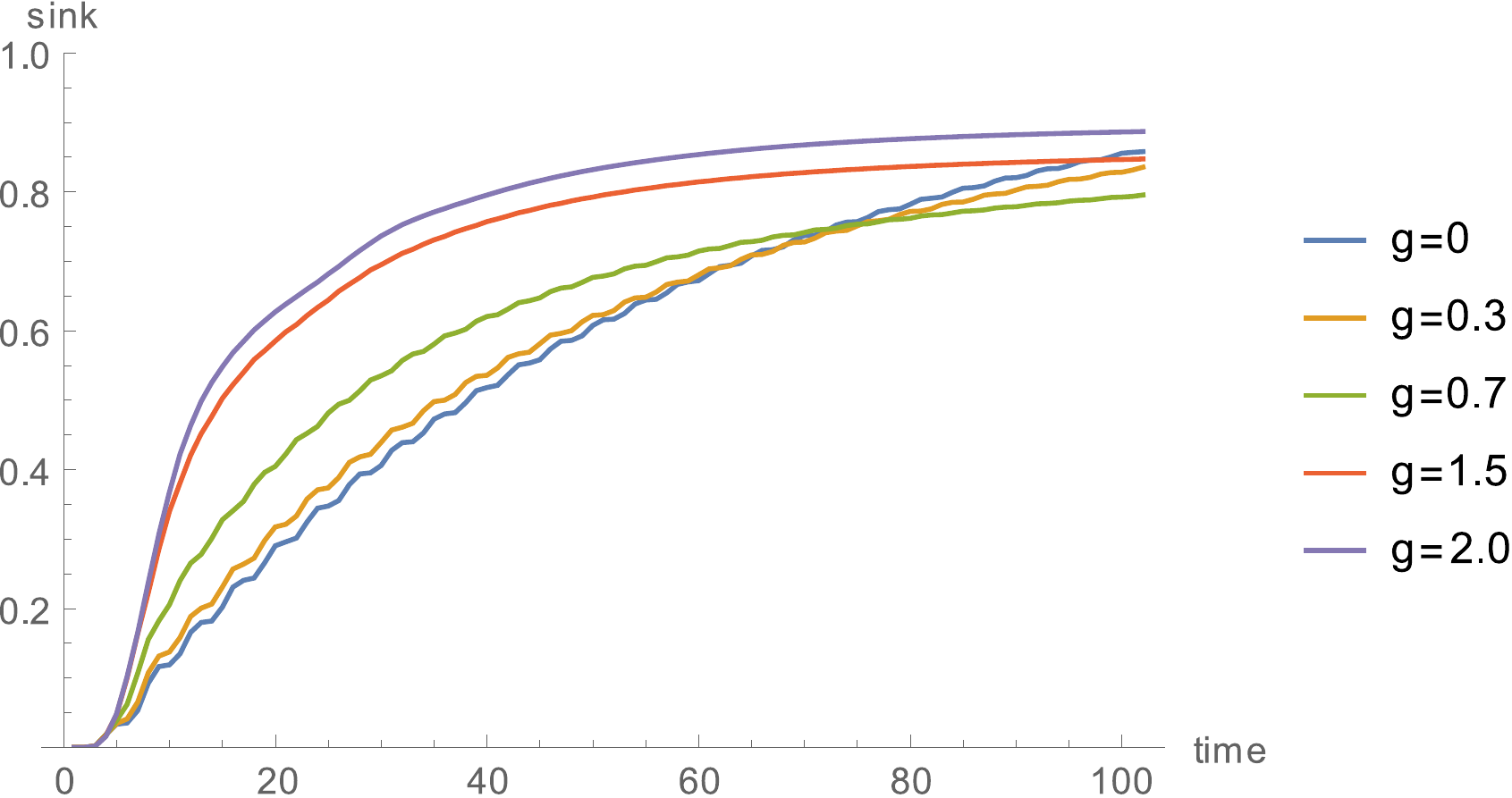} & & \includegraphics[width=0.45\textwidth]{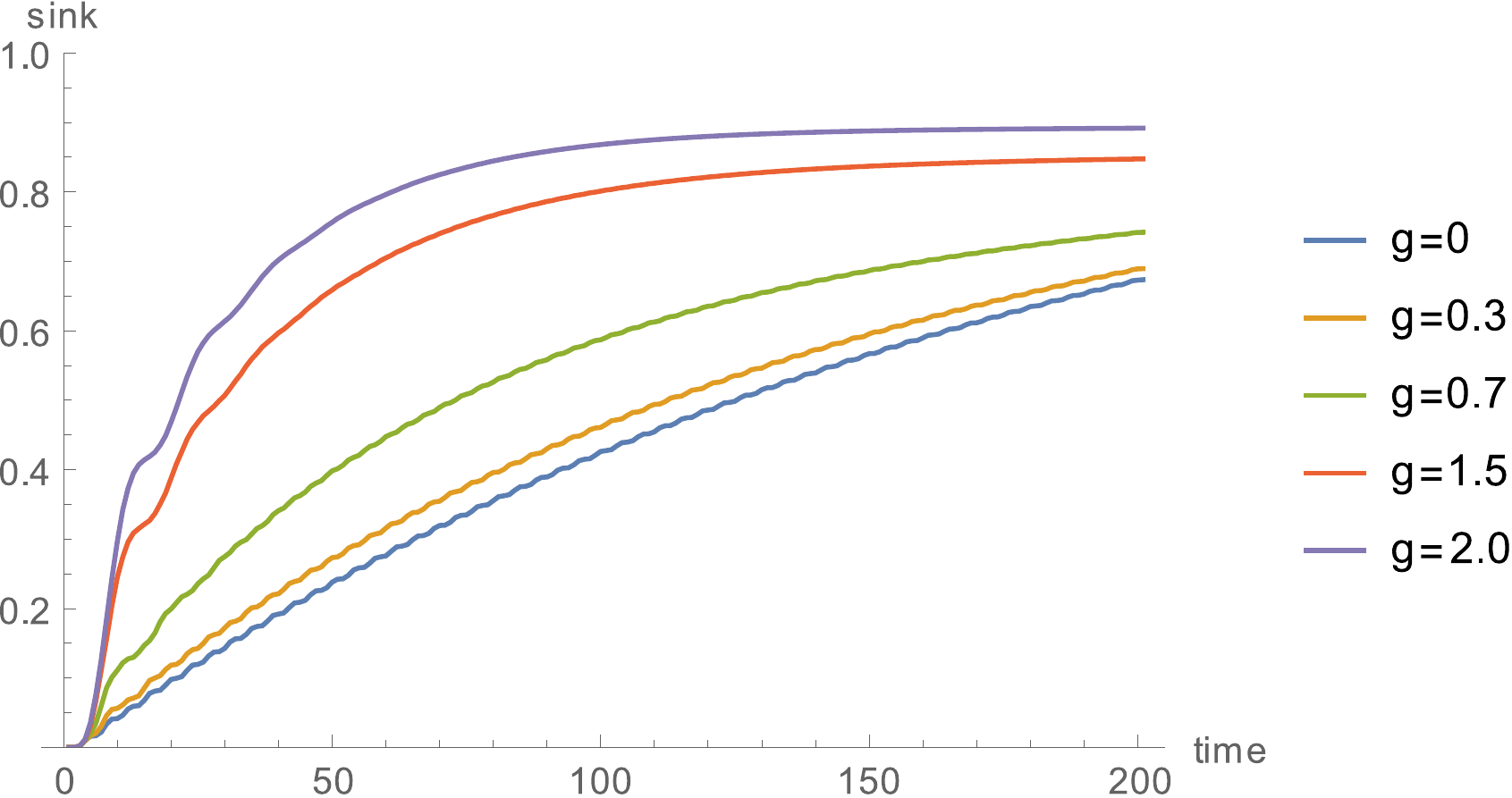} \\
$out$ = 2.0 & & $out$ = 4.0 \\
\end{tabular}
\vspace{0.2cm}
 \caption{ \label{fig:2-udat-time0} Evolution of the sink state over time. Two atoms, no input (initial state has a photon in the first cavity). $k$ = 0.2, $\mu$ = 0.8. }
\end{figure}

\pagebreak

\section{Conclusions}
We have shown the efficiency of the qubit representation of a JCH model with the Lindblad-like form of dephasing, and its relation with the simple unitary model of dephasing.

Quantum bottleneck effect was explained and reproduced for both input and output rates. It was shown that, depending on the model layout, the optimal input rate could mismatch the optimal output rate. The dependency of the bottleneck effect on the in-model interaction rates was shown.

Two models of dephasing showed similar results in the long term, when the numeric experiments were run long enough for the sink to reach a value close to $1$. For both models non-zero dephasing did not give a positive effect on the conductivity in the case of optimal output rate (when the quantum bottleneck effect is not visible, but gave a large positive effect in some cases when the output rate was not optimal and when the conductivity is capped by the quantum bottleneck effect.

It was also shown that short-term effects differ from the long-term effects, and while some settings are optimal for reaching a low target sink value (for example $0.3$), they could be far from optimal for reaching a close to $1$ target value.

\end{document}